%% file: 17hcc.tex
\documentclass{aa}

\usepackage{txfonts}
\usepackage{rotating}
\usepackage{color}
\usepackage{xcolor}
\usepackage{orcidlink}
\usepackage{float}
\usepackage{longtable}
\usepackage{lscape}

\usepackage[utf8]{inputenc}

\usepackage{ulem}
\usepackage{endnotes}

\let\footnote=\endnote 

\usepackage{amsmath} 

\makeatother

\usepackage{natbib}
\bibpunct{(}{)}{;}{a}{}{,} 
 
\newcommand{\kms}{km\,s$^{-1}$}

\begin{document} 

   \title{A long life of excess: The interacting transient SN~2017hcc}

   \author{S. Moran \orcidlink{0000-0001-5221-0243}\inst{1}
          \and
          M. Fraser \orcidlink{0000-0003-2191-1674}\inst{2} 
          \and
          R. Kotak \orcidlink{0000-0001-5455-3653}\inst{1}
          \and
          A. Pastorello\inst{3}
          \and 
          S. Benetti \orcidlink{0000-0002-3256-0016}\inst{3}
          \and
          S.J. Brennan\inst{2}
          \and
          C.P. Guti\'errez\inst{1, 4}
          \and
          E. Kankare\inst{1}
          \and
          H. Kuncarayakti \orcidlink{0000-0002-1132-1366}\inst{1}
          \and
          S. Mattila\inst{1,5}
          \and
          T.M. Reynolds\inst{1, 6, 7}
          \and
          J.P. Anderson \orcidlink{0000-0003-0227-3451}\inst{8}
          \and P.J. Brown\inst{9}, 
          S. Campana\inst{10} \and K.C. Chambers\inst{11}
          \and T.-W. Chen \orcidlink{0000-0002-1066-6098}\inst{12} \and M. Della Valle \orcidlink{0000-0003-3142-5020}\inst{13,14,15} \and M. Dennefeld\inst{16} \and N. Elias-Rosa\inst{3,17} \and L. Galbany \orcidlink{0000-0002-1296-6887}\inst{17,18} \and F.J. Galindo-Guil \orcidlink{0000-0003-4776-9098}\inst{19} \and M. Gromadzki\orcidlink{0000-0002-1650-1518}\inst{20} \and D. Hiramatsu \orcidlink{0000-0002-1125-9187}\inst{21,22,23,24} \and C. Inserra \orcidlink{0000-0002-3968-4409}\inst{25} \and G. Leloudas \orcidlink{0000-0002-8597-0756}\inst{26} \and T.E. M\"uller-Bravo \orcidlink{0000-0003-3939-7167}\inst{17} \and M. Nicholl\inst{27} \and A. Reguitti \orcidlink{0000-0003-4254-2724}\inst{28,29,3} \and M. Shahbandeh \orcidlink{0000-0002-9301-5302}\inst{30} \and S.J. Smartt\inst{31} \and L. Tartaglia \orcidlink{0000-0003-3433-1492}\inst{3} \and D.R. Young \orcidlink{0000-0002-1229-2499}\inst{31}
          }

          \institute{Department of Physics and Astronomy, University of Turku, Vesilinnantie 5, FI-20500, Finland
          \and
          School of Physics, University College Dublin, Belfield, Dublin 4, Ireland
          \and INAF – Osservatorio Astronomico di Padova, Vicolo dell'Osservatorio 5, I-35122 Padova, Italy
          \and Finnish Centre for Astronomy with ESO (FINCA), FI-20014 University of Turku, Finland 
          \and School of Sciences, European University Cyprus, Diogenes street, Engomi, 1516 Nicosia, Cyprus
          \and Cosmic Dawn Center (DAWN), Denmark  
          \and Niels Bohr Institute, University of Copenhagen, Jagtvej 128, 2200 København N, Denmark
          \and European Southern Observatory, Alonso de C\'ordova 3107, Casilla 19, Santiago, Chile 
          \and George P. and Cynthia Woods Mitchell Institute for Fundamental Physics \& Astronomy, Texas A. \& M. University, Department of Physics and Astronomy, 4242 TAMU, College Station, TX 77843, USA 
          \and INAF – Osservatorio astronomico di Brera, Via Bianchi 46, I-23807, Merate (LC), Italy
          \and Institute for Astronomy, University of Hawaii, 2680 Woodlawn Drive, Honolulu, HI 96822, USA  
          \and The Oskar Klein Centre, Department of Astronomy, Stockholm University, AlbaNova, SE-10691 Stockholm, Sweden
          \and INAF – Capodimonte Astronomical Observatory, Salita Moiariello 16, I-80131 Napoli, Italy
          \and INFN-Napoli, Strada Comunale Cinthia, I-80126 Napoli, Italy
          \and ICRANet, Piazza della Repubblica 10, I-65122 Pescara, Italy
          \and Institut d'Astrophysique de Paris (IAP), CNRS \& Sorbonne Universite, 75014 Paris, France 
          \and Institute of Space Sciences (ICE, CSIC), Campus UAB, Carrer de Can Magrans s/n, E-08193 Barcelona, Spain 
          \and Institut d'Estudis Espacials de Catalunya (IEEC), E-08034 Barcelona, Spain
          \and Centro de Estudios de F\'isica del Cosmos de Arag\'on (CEFCA), Plaza San Juan 1, 44001 Teruel, Spain
          \and Astronomical Observatory, University of Warsaw, Al. Ujazdowskie 4, 00-478 Warszawa, Poland
          \and Center for Astrophysics \textbar{} Harvard \& Smithsonian, 60 Garden Street, Cambridge, MA 02138-1516, USA
          \and The NSF AI Institute for Artificial Intelligence and Fundamental Interactions
          \and Las Cumbres Observatory, 6740 Cortona Drive, Suite 102, Goleta, CA 93117-5575, USA
          \and Department of Physics, University of California, Santa Barbara, CA 93106-9530, USA
          \and School of Physics and Astronomy, Cardiff University, Queens Buildings, The Parade, Cardiff CF24 3AA, UK 
          \and DTU Space, National Space Institute, Technical University of Denmark, Elektrovej 327, 2800 Kgs. Lyngby, Denmark
          \and Birmingham Institute for Gravitational Wave Astronomy and School of Physics and Astronomy, University of Birmingham, Birmingham B15 2TT, UK
          \and Instituto de Astrof\'{i}sica, Departamento de Ciencias F\'{i}sicas – Universidad Andres Bello, Avda. Rep\'{u}blica 252, 8320000, Santiago, Chile
          \and Millennium Institute of Astrophysics, Nuncio Monsenor S\'{o}tero Sanz 100, Providencia, 8320000, Santiago, Chile
          \and Department of Physics, Florida State University, 77 Chieftan Way, Tallahassee, FL 32306, USA
          \and Astrophysics Research Centre, School of Maths and Physics, Queen's University Belfast, Belfast BT7 1NN, UK 
          }

   \date{Received ?; accepted ?}

    \abstract{
    In this study we present the results of a five-year follow-up campaign of the long-lived type IIn supernova SN~2017hcc, found in a spiral dwarf host of near-solar metallicity. The long rise time (57 $\pm$ 2 days, ATLAS $o$ band) and high luminosity (peaking at $-$20.78 $\pm$ 0.01 mag in the ATLAS $o$ band) point towards an interaction of massive ejecta with massive and dense circumstellar material (CSM). The evolution of SN~2017hcc is slow, both spectroscopically and photometrically, reminiscent of the long-lived type IIn, SN~2010jl. An infrared (IR) excess was apparent soon after the peak, and blueshifts were noticeable in the Balmer lines starting from a few hundred days, but appeared to be fading by around +1200\,d. We posit that an IR light echo from pre-existing dust dominates at early times, with some possible condensation of new dust grains occurring at epochs >$\sim$+800\,d.}
    

   \maketitle
%
\section{Introduction}

Supernovae (SNe) are traditionally grouped into two types: type I (hydrogen-poor) and type II (hydrogen-rich) \citep{Minkowski1941}. There are subdivisions within both of these classes of SNe based on criteria such as spectral lines and light curve behaviour \citep{Gal-Yam2017}.
The type IIn designation was first introduced by \cite{Schlegel1990} to describe SNe with narrow (FWHM $\lesssim1000$~\kms) hydrogen emission lines in their spectra. 
These narrow lines arise from dense circumstellar material ionised by 
an ongoing shock interaction \citep[e.g.][]{Chugai1997}. 
Most SNe IIn are expected to have progenitor stars with a high mass loss rate, providing the necessary circumstellar material (CSM) for later shock interactions post-explosion, and they have often been associated with luminous blue variable-like (LBV-like) stars \citep{Kiewe12, Smith17}.

Type IIn SNe are relatively rare, constituting about 9\% of all core-collapse SNe in the local Universe (\citealt{Li11, Cappellaro2015}); however, these estimates are uncertain, with \cite{2013Eldridge} finding a much lower rate of 2.4\%, for example. 
Type IIn SNe show tremendous diversity in their light curves and spectra \citep[e.g.][]{Kiewe12,Nyholm2020,Fraser2020}; this diversity likely reflects a large spread of CSM masses and density profiles, geometric configurations, wind velocities, chemical compositions, as well as properties of the progenitors and their environments. Some of the most extreme exemplars are characterised by long-lasting, slowly evolving light curves. SN~2010jl is an archetypal long-lived type IIn event, reaching an absolute magnitude of $\sim -$19.9 mag in the $V$ band and remaining bright for several years \citep{Stoll2011,Ofek2019}.

\begin{figure}
\includegraphics[width=1\linewidth]{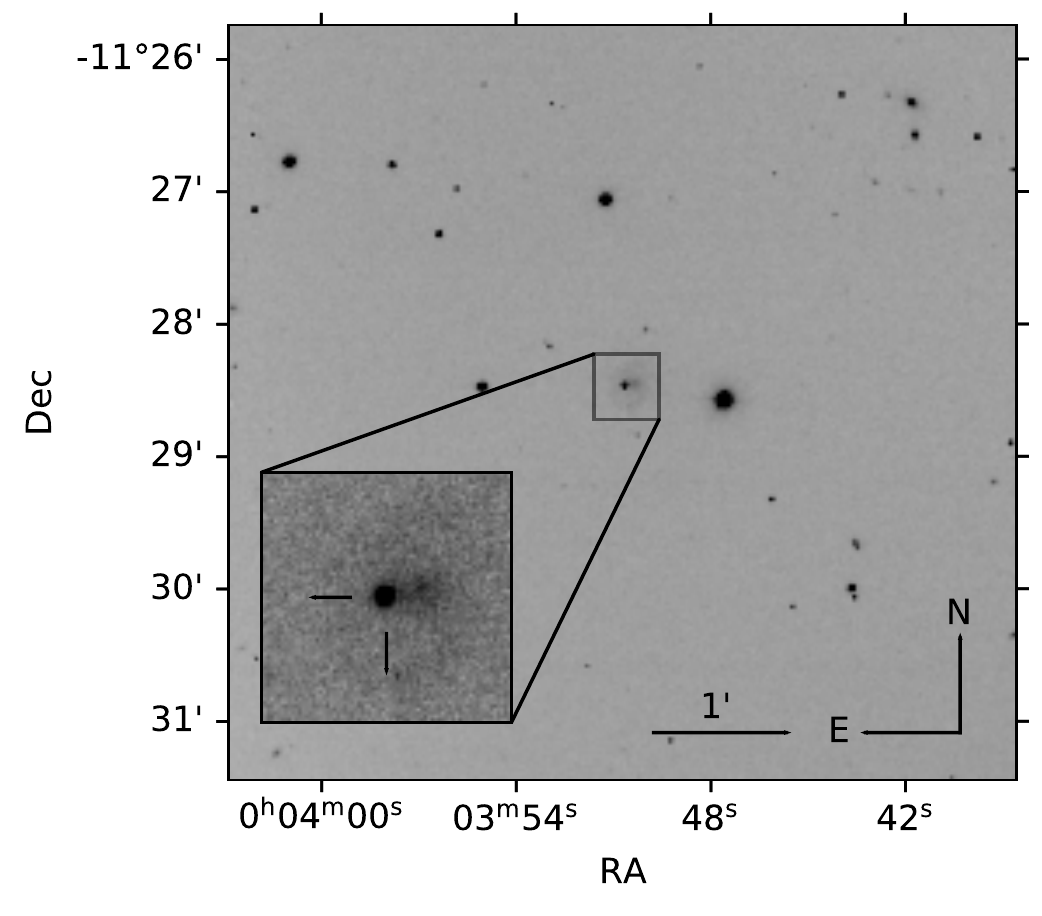}
\caption{SN~2017hcc ($\alpha=00^{\mathrm{h}}03^{\mathrm{m}}50^{\mathrm{s}}.58$, $\delta =-11^{\circ}28'28''.78$ J2000), based on a $V$ band image taken with NOT+ALFOSC on 2018 Sept 30 (+365\,d). The inset shows the region around the SN in more detail, with the faint nucleus of the host just visible to the west of the SN.} 
\label{fig:finder}
\end{figure}
In this study we report on the long-lived, slowly evolving type IIn SN~2017hcc. As previously mentioned, SNe IIn are a heterogeneous group, so photometry and spectroscopy covering key epochs are a pre-requisite to deciphering the pre-explosion history of the progenitor. 
We present results from a comprehensive ultraviolet (UV) to mid-infrared (MIR) follow-up campaign conducted over a five year period.

SN~2017hcc was discovered in the course of the Asteroid Terrestrial-impact Last Alert System (ATLAS) survey \citep{Tonr18, SmithKW2020} on 2017 Oct 2 \citep[internal survey designation ATLAS17lsn,][]{Tonr17}, and was classified by the All-Sky Automated Survey for SuperNovae (ASAS-SN, \citealt{Shappee2014, Koch17}) as a type IIn SN \citep{Dong17}. 
The SN is located at $\alpha=00^{\mathrm{h}}03^{\mathrm{m}}50^{\mathrm{s}}.58$, $\delta =-11^{\circ}28'28''.78$ J2000 (see Fig. \ref{fig:finder}).

A number of surveys independently discovered SN~2017hcc. The {\it Gaia} Science Alerts \citep{Hodgkin2021} project designated it Gaia17dcj; the Panoramic Survey Telescope and Rapid Response System (Pan-STARRS1; \citealp{Huber15}) project called it PS17fra; and the Zwicky Transient Facility (ZTF; \citealp{Bellm19}) called it ZTF18abtmgfn. 

Based on the measured wavelength of H$\alpha$ in the +88\,d Nordic Optical Telescope (NOT) + ALFOSC {\it gr17} spectrum, we adopted a heliocentric redshift of {\it z} = 0.0168 $\pm$ 0.0001 for SN~2017hcc, which is in line with the measurement obtained by \cite{Prieto2017}. 
For a Hubble parameter of 70 km~s$^{-1}$~Mpc$^{-1}$, taken as a compromise between inconsistent measured values \citep{2019Freedman, 2021DiValentino}, we calculated the luminosity distance to be 72.9 $\pm$ 0.4 Mpc and the distance modulus to be $\mu=34.31 \pm 0.01$~mag. Unfortunately, there was no progenitor candidate observed for SN~2017hcc due to its distance. 
The explosion epoch has been set as MJD 58\,027.4 (2017 Oct 1 09:36:00 UTC, Sect. \ref{sect:phot_evo}). All phases reported in this paper are with respect to this date.

We take the foreground extinction towards SN~2017hcc to be $A_V=0.091$ mag from \cite{Schl11} (via the NASA Extragalactic Database, NED\footnote{\url{http://ned.ipac.caltech.edu/}}), assuming an $R_V$ of 3.1 \citep{Schultz1975}. 
The host galaxy of SN~2017hcc appears to be a dwarf spiral (Fig. \ref{fig:finder}). It has an apparent magnitude of r$_{Kron}$ = 17.8 mag in the Pan-STARRS1 survey. For a distance of 72.9\,Mpc, this implies an absolute magnitude of $-$16.6 mag, taking Milky Way extinction into account, which is suggestive of a low mass host \citep{Pskovskii1965}. 

The equivalent width of narrow sodium lines has frequently been used as a proxy for extinction (e.g. \citealt{Poznanski2012}). Whilst there are several concerns regarding the reliability of such methods (e.g. \citealt{Poznanski2011}), in the case of SN~2017hcc the presence of any significant sodium absorption in any of our spectra is unclear, which is at least qualitatively consistent with very little dust in the host line of sight. This is corroborated by \citet{Smith2020}, who inferred an upper limit of E($B-V$) = 0.016 for the host extinction using echelle spectra. Additionally, the spectra of SN~2017hcc are quite blue when compared to other type IIn SNe, which gives a further indication of low extinction. As such, we regard extinction in the host galaxy to be negligible.

The paper is organised as follows: in Section \ref{sect:obs_data}, we discuss the observational data acquisition and reduction. These data comprise ultraviolet, optical and near-infrared (NIR) imaging along with optical and NIR spectroscopy. The photometric and spectroscopic evolution is outlined in Sections \ref{sect:phot_evo} and \ref{sect:spec_evo}, respectively. 
In Section \ref{sect:metallicity} we describe the calculation of the metallicity of the environment. 
Section \ref{sect:disc_conc} contains a summary and our conclusions. All photometric data and logs of the optical and NIR spectroscopy are listed in tables in the appendix. We note that, alongside the previous works by \cite{Prieto2017}, \cite{Kumar2019}, and \cite{Smith2020}, a detailed study by \cite{Chandra2022} concentrating on the X-ray, radio, and infrared (IR) properties of SN~2017hcc was posted to the arXiv preprint server whilst this work was under review.

\section{Data acquisition and reduction}
\label{sect:obs_data}

\subsection{Optical imaging}
We have made use of optical imaging data from a number of telescopes and instruments for SN~2017hcc: the Las Cumbres Observatory (LCO) network of 1m telescopes, the NOT + ALFOSC, the Liverpool Telescope + IO:O, the 0.9m Asiago Schmidt telescope + KAF-16803 CCD. 
We make use of observations taken in Sloan {\it ugriz} filters, as well as Johnson-Cousins {\it B} and {\it V}. In addition, photometry was obtained as part of routine operations of the ATLAS, Pan-STARRS1, and ASAS-SN surveys and the {\it Gaia} Alerts project. The Pan-STARRS1 imaging was taken in the {\it w} and {\it i} filters, whilst for ATLAS cyan ({\it c}) and orange ({\it o}) filters were used, where {\it c} is comparable to {\it g+r} whilst {\it o} is comparable to {\it r+i}. All {\it Gaia} photometry was taken in the {\it Gaia G} filter, a wide band filter that covers approximately {\it g+r+i}. The ASAS-SN photometry was taken in the {\it V} and {\it g} filters. Though there is ZTF photometry available for SN~2017hcc, we have not made use of it, since a template issue rendered the results unreliable. 

We reduced all optical images in a similar fashion. The images were bias and overscan subtracted before being flat-fielded and trimmed.  
For the NOT+ALFOSC images, which were largely obtained via the NUTS and NUTS2 programmes\footnote{\url{https://nuts.sn.ie/}}, 
and the Asiago images, we performed these steps with dedicated {\sc foscgui} pipelines\footnote{\url{https://sngroup.oapd.inaf.it/foscgui.html}} for data reduction developed by E. Cappellaro. 
The data from the Liverpool Telescope were reduced automatically by the IO:O pipeline\footnote{\url{https://telescope.livjm.ac.uk/TelInst/Pipelines/\#ioo}}, whilst the LCO data were reduced automatically by the {\sc banzai} pipeline\footnote{\url{https://github.com/LCOGT/banzai}}. The ATLAS, ASAS-SN, and Pan-STARRS1 data were reduced automatically by their respective pipelines \citep{SmithKW2020,Koch17,Magnier2020} with Point-Spread Function (PSF) fitting being used to obtain photometry.

From the start of the third observing season ($\gtrsim$+600\,d), the increasing contribution of the host galaxy became apparent. Since SN~2017hcc was still visible in our final observations, we were unable to obtain fresh SN-free template images in order to perform difference imaging, though shallow Pan-STARRS1 templates did exist for the field. As a result of this limitation, we used the {\sc autophot}\footnote{\url{https://github.com/Astro-Sean/autophot}} code \citep{Brennan2022} to measure the brightness of SN~2017hcc with PSF-fitting photometry. 

We calibrated the zeropoints for the {\it B} and {\it V} bands against the APASS catalogue, whilst we calibrated those of the {\it g}, {\it r}, {\it i} and {\it z} bands against the Pan-STARRS1 catalogue. Full details of all optical photometric measurements determined with {\sc autophot} are given in Table \ref{tab:optphot}.
\subsection{Ultraviolet and X-ray imaging}
SN~2017hcc was also observed by the UltraViolet Optical Telescope (UVOT; \citealp{Roming2005}) on the Neil Gehrels {\it Swift} Observatory. {\it Swift}/UVOT data have been previously published by \citet{Prieto2017} and \citet{Kumar2019}. We use photometry created by the pipeline for the {\it Swift} Optical Ultraviolet Supernova Archive (SOUSA; \citealp{Brown2014}), updated with the zeropoints of \citet{Breeveld2010} and the updated sensitivity correction from CALDB 20200925. We did not perform host galaxy template subtraction. Full details of all {\it Swift} photometric measurements are given in Table \ref{table:swift}. 

Together with UVOT, {\it Swift} has a co-pointing X-ray instrument: the X-Ray Telescope (XRT, \citealt{Burrows2005}). We summed the 13 observations taken during the first 45 days, ending with 28.6\,ks of observing time. We do not detect any source at the location of SN~2017hcc. The $3\,\sigma$ upper limit on the 0.3-10\,keV count rate is $8.4\times10^{-4}$ c s$^{-1}$. Assuming a power law spectral model with a photon index $\Gamma=2$ and a Galactic column density of $2.9\times 10^{20}$ cm$^{-2}$ \citep{Willingale2013}, one can translate this into an upper limit on the 0.3-10\,keV unabsorbed flux of $<3.7\times 10^{-14}$ erg cm$^{-2}$ s$^{-1}$, or on the X-ray luminosity of $<2.5\times 10^{40}$ erg s$^{-1}$.

\subsection{Infrared imaging}

We used NOT+NOTCam, as part of the NUTS and NUTS2 programmes, and the New Technology Telescope (NTT) with the SOFI instrument for NIR imaging of SN~2017hcc, as part of the ePESSTO programme\footnote{\url{https://www.pessto.org/}} \citep{Smar15}. 
We used the NOTCam QUICKLOOK\footnote{\url{http://www.not.iac.es/instruments/notcam/guide/observe.html}} v2.5 reduction package, which is a set of IRAF scripts, with a few functional modifications (e.g. to increase the FOV of the reduced image), to reduce the NOTCam images. Bright and faint sky flats were taken and these were used to create a differential master flat. The flats were checked to make sure the count levels were linear. We masked out bad pixels (cold, hot, zero) during reduction and we used a model to account for the significant optical distortion of the NOTCam wide-field camera. We accounted for the dark current through the use of differential flats in the reduction process.

For SOFI reductions, we used the \textsc{pessto pipeline}\footnote{\url{https://github.com/svalenti/pessto}} \citep{Smar15}. The raw frames were corrected for array crosstalk, and a sky image was created which was subtracted from the individual frames. The frames were flat-fielded with differential dome flats created by taking flats with the dome lamp on and off. An illumination correction was applied to account for the difference between sky and dome illumination patterns. The pipeline uses SExtractor to detect objects in the dithered images. The dithered images are aligned and combined using IRAF tasks based on a transformation between corresponding objects in the images.

As in the case of the optical bands, we used the {\sc autophot} code to measure the brightness of SN~2017hcc in NIR bands with PSF-fitting photometry. The zero points were calculated using the Two Micron All-Sky Survey (2MASS) catalogue\footnote{\url{https://irsa.ipac.caltech.edu/Missions/2mass.html}}. As the 2MASS catalogue does not have $Ks$ band photometry, we used the $K$ band measurements of the catalogue as a substitute. Full details of all NIR photometric measurements determined with {\sc autophot} are given in Table \ref{table:NIR}.

\begin{figure*}
\includegraphics[width=1\linewidth]{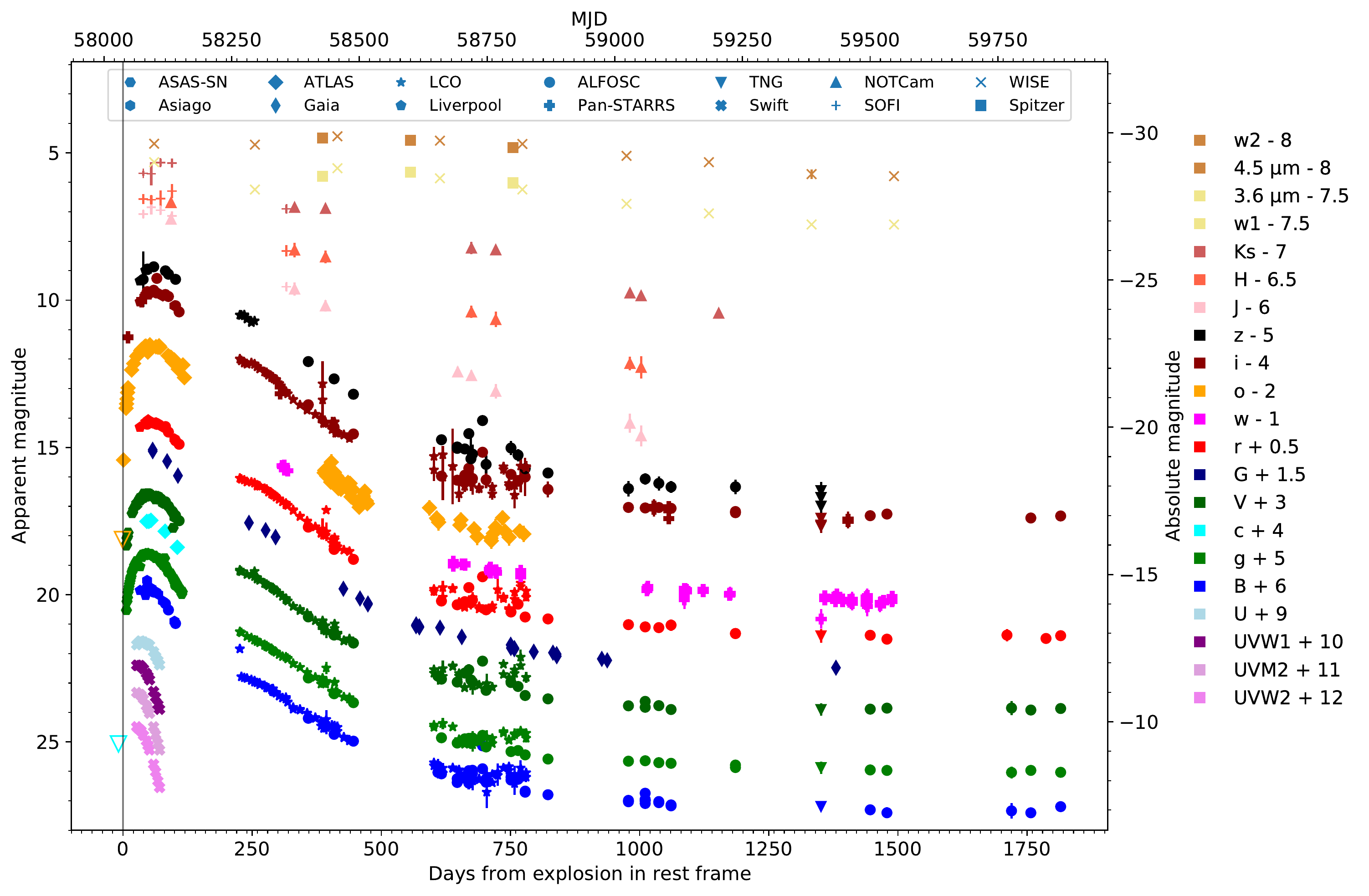}
  \caption{Full UV to MIR light curve of SN~2017hcc. The ATLAS points have been averaged to within 0.3 MJD. The ATLAS upper limits have been marked with inverted open triangles. A vertical line has been placed at the explosion epoch to guide the eye.} 
  \label{fig:lightcurve}
\end{figure*}

The W1 (3.4$\mu$m) and W2 (4.6$\mu$m) NEOWISE \citep{Mainzer2011, Mainzer2014} photometry was taken from the public NEOWISE-R Single Exposure (L1b) Source Table\footnote{\url{https://irsa.ipac.caltech.edu/cgi-bin/Gator/nph-scan?mission=irsa&submit=Select&projshort=WISE}} and can be found in Table \ref{table:WISE}. We adopted the median value of the individual measurements at each observing visit as the magnitude after removing data flagged as poor quality, or separated by more than 2" from the SN location. For the uncertainty, we adopted the standard error of mean of the individual measurements after 3 sigma-clipping outliers, with an additional uncertainty associated with the photometric calibration added in quadrature. We also include three epochs (393, 566, and 768 d) of Spitzer imaging at 3.6$\mu$m and 4.5$\mu$m in our analysis, adopting the photometric measurements of \cite{Szalai2021}. 

\subsection{Optical spectroscopy}

The bulk of the spectra were taken at the NOT using the ALFOSC instrument (23 out of the total of 52) as part of the NUTS and NUTS2 programmes, and at the NTT with the EFOSC2 instrument (13 of 52) as part of the ePESSTO and ePESSTO+ programmes. Four spectra were taken at the Telescopio Nazionale Galileo (TNG) with the DOLORES instrument and two were taken at the Gran Telescopio Canarias (GTC) with the OSIRIS instrument. One was taken at the Very Large Telescope (VLT) using the Multi Unit Spectroscopic Explorer (MUSE) instrument. Two using the FLOYDS instrument at the FTS telescope as part of the LCO network of telescopes. One at the Asiago 1.82m telescope using the AFOSC instrument and five using the Asiago 1.22m and the B\&C instrument.
The complete log of optical spectral observations can be found in Table \ref{tab:spec}.

We reduced the NTT+EFOSC2 spectra using the \textsc{pessto pipeline} \citep{Smar15}. 
The raw two-dimensional spectra were overscan and bias subtracted, and divided by a normalised flat field before masking cosmic rays. The spectra were wavelength calibrated using an arc lamp to give the dispersion solution, together with an additional offset measured from bright sky emission lines in order to account for flexure within the instrument. Flux calibration was achieved using a master sensitivity function created from spectra of between three and seven spectro-photometric standard stars, observed at similar times to the science spectra. 
We combined {\it gr11} and {\it gr16} spectra taken on the same night into individual spectra using the \texttt{scombine} IRAF command.

We reduced the NOT+ALFOSC, GTC+OSIRIS and TNG+DOLORES spectra in an analogous fashion. For the NOT+ALFOSC spectra, with the exception of the sole {\it gr17} spectrum, we performed the reductions using the {\sc foscgui} pipeline. In the case of the NOT+ALFOSC {\it gr17} spectrum, the GTC+OSIRIS spectrum and the TNG+DOLORES spectra, we performed the reduction manually, using IRAF tasks. The DOLORES spectra have been shifted slightly (by a few {\AA}) based on the position of the 6300.31 {\AA} skyline. 
We extracted the spectrum from already processed data cube taken with the MUSE instrument on VLT as part of the All-weather MUse Supernova Integral-field of Nearby Galaxies (AMUSING; \citealt{Galbany2016}). The extraction was performed using the {\sc qFitsView}\footnote{\url{https://www.mpe.mpg.de/~ott/dpuser/qfitsview.html}} software package, by centring an annulus of radius one arc second on the location of SN~2017hcc. The FTS+FLOYDS spectra of which we make use were automatically reduced by the FLOYDS Pipeline\footnote{\url{https://github.com/griffin-h/floyds_pipeline}}\footnote{\url{https://lco.global/documentation/data/floyds-pipeline/}} at the LCO headquarters following the observation nights.

\subsection{Infrared spectroscopy}
Near-infrared spectra were taken with NTT+SOFI using an ABBA pattern of nodding along the slit. We reduced these data using the {\sc pessto pipeline} \citep{Smar15}. The raw two-dimensional spectra were calibrated using an arc lamp and the frames were corrected for array crosstalk and flat-fielded. 
Frames taken at position A were subtracted from those taken at position B (and vice-versa) to remove the varying sky background. Finally, the individual exposures were shifted and combined, before a one-dimensional spectrum was optimally extracted. A spectrum of a telluric standard was taken immediately after each science spectrum at a similar airmass, and this was used to correct for telluric absorption. The telluric standard was also used to achieve a flux calibration for each spectrum. We also obtained spectra from the NASA Infrared Telescope Facility (IRTF) using the SpeX instrument. These data were reduced using the IDL-based \texttt{Spextool} \citep{Cushing2004} in an analogous fashion. The complete log of NIR spectral observations can be found in Table \ref{tab:NIR_spec}.

\section{Photometric evolution}
\label{sect:phot_evo}
\subsection{Light curve and colour}

\begin{figure}
\includegraphics[width=\linewidth]{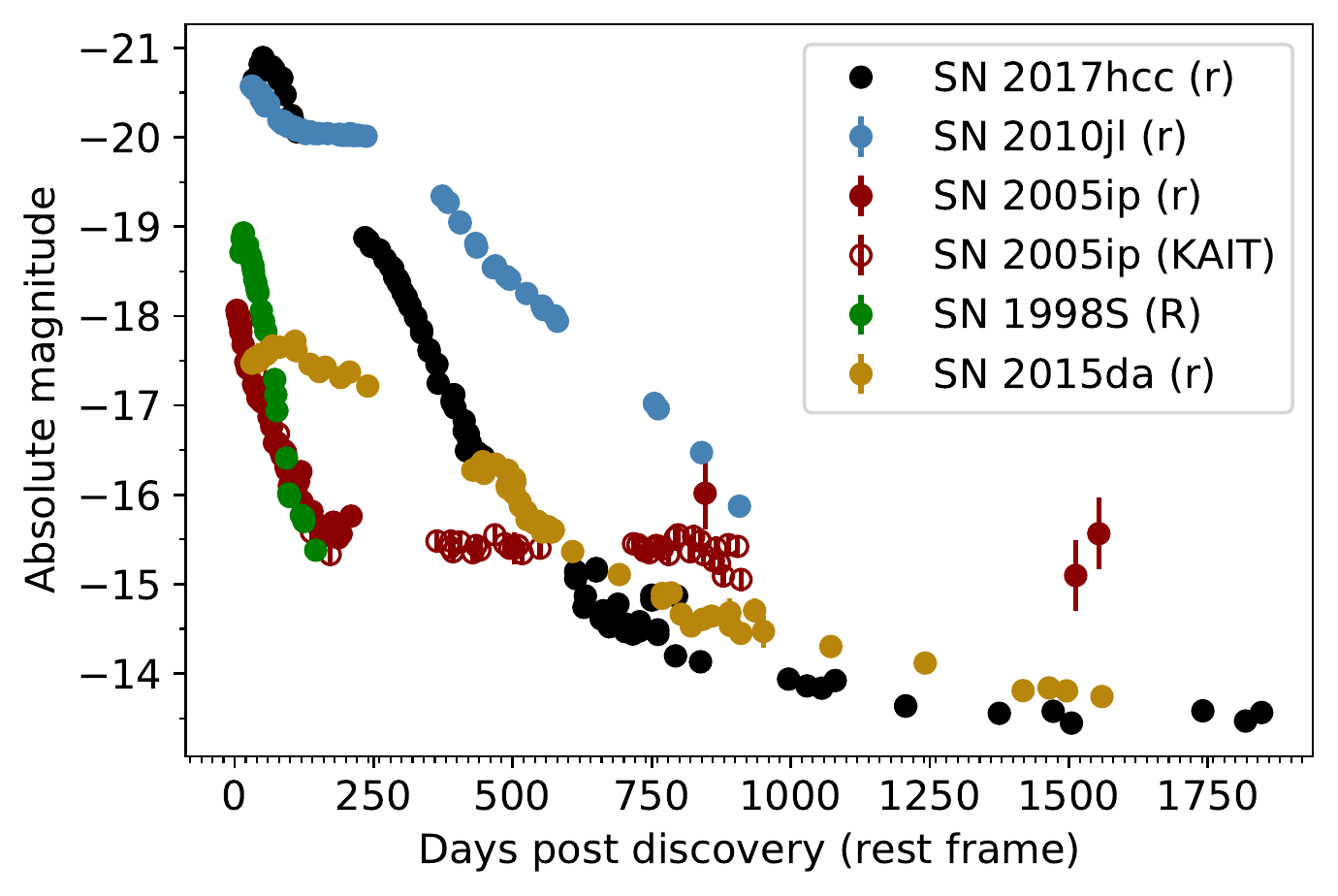}
  \caption{Comparison of the $r$ band light curve of SN~2017hcc with that of SN~2010jl \citep{Fransson2014} and SN~2005ip \citep{Stritzinger2012}, SN~2015da \citep{Tartaglia2020} along with the $R$ band photometry of SN~1998S \citep{Fassia2000}. The open circles for SN~2005ip are unfiltered KAIT data \citep{Smith2009}, which are taken to be roughly equivalent to $R$ band. The error bars, where not visible, are smaller than the points. All light curves have been corrected for foreground extinction and time dilation. Phases are given against discovery date. Following \citep{Tartaglia2020},
  we do not account for the host extinction of SN~2015da, but note that it could be significant ($A_R$ = 2-3\, mag).}

  \label{fig:lightcurve_comp}
\end{figure}

\begin{figure}
\includegraphics[width=\linewidth]{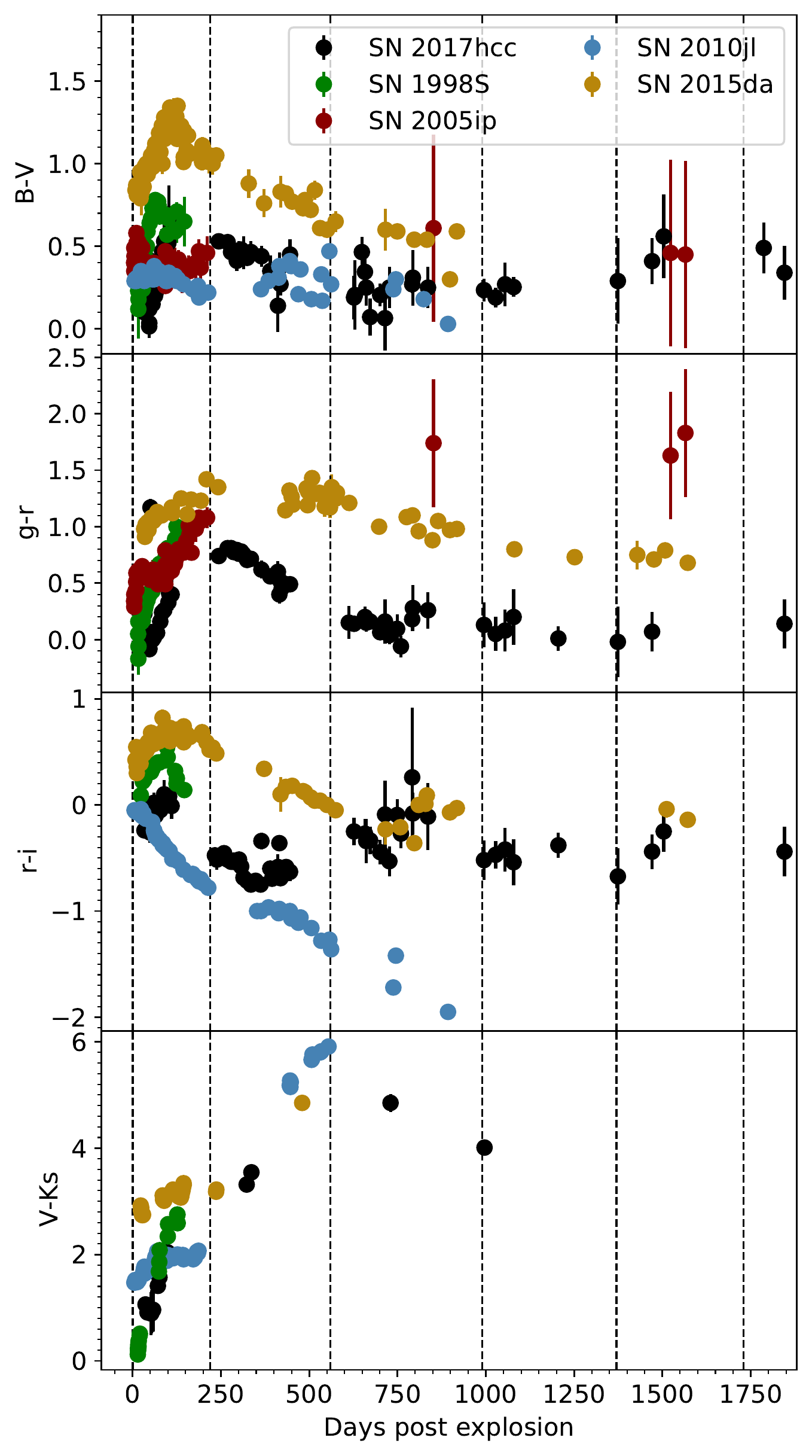}
  \caption{From top to bottom, $B-V$, $g-r$, $r-i$, and $V-K_\mathrm{s}$ colour curves for SN~2017hcc, compared to a selection of SNe IIn. The photometry has been averaged to one MJD (with errors added in quadrature), with colours then having been calculated with observations within 1one day of each other in the relevant bands, except in the case of the $V-K_\mathrm{s}$ observations where a larger window of seven days has been allowed, due to the paucity of observations. The errors bars, when not visible, are smaller than the points. Phases are given against discovery date, except in the case of SN~2017hcc, where the explosion epoch is used. No extinction corrections have been applied to the colours. In the second panel the SN~1998S data are actually $V-R$ rather than $g-r$ and in the third panel both the SN~1998S data and the SN~2015da data are $R-I$ rather than $r-i$. In the final panel the SN~2010jl and SN~1998S data are actually $V-K$ rather than $V-K_\mathrm{s}$. The dashed vertical lines mark the beginning of seasons one, two, three and four, five and six, respectively.} 
  \label{fig:col_curves}
\end{figure}

SN~2017hcc is an extremely long-lived SN, characterised by a slowly evolving and luminous light curve (see Fig. \ref{fig:lightcurve}). 
The region of the sky where the SN exploded had been well-monitored by the ATLAS survey in the period immediately prior to the discovery epoch. 
Taking a midpoint between the first detection (MJD 58\,028.4, 17.44 mag, ATLAS-$o$) and last non-detection (MJD 58\,026.4, ATLAS-$o$ $>19.04$ mag)\footnote{The limiting magnitude determined by forced photometry on ATLAS images at the SN location is slightly deeper in the ATLAS-$o$ (>19.69 mag). We adopt the more conservative and shallower limit here, although either limit would imply an explosion epoch some time during the two day period between this epoch and discovery.}, we estimate an explosion epoch of MJD 58027.4 $\pm$ 1.0 for SN~2017hcc.
SN~2017hcc has a slow rise to maximum, as can be seen in Fig. \ref{fig:lightcurve}. 
To determine the peak of the light curve, we fitted a third order polynomial to the ATLAS $o$ band light curve. We found the $o$ band maximum to be at MJD 58084.1 at absolute magnitude $-$20.78 $\pm$ 0.01 (apparent magnitude 13.53 $\pm$ 0.01), which gives a rise time of 57 $\pm$ 2 days. 

After maximum, the light curve begins a slow decline in the optical bands. In contrast to the optical, the UV light curve of SN~2017hcc declines from the first epoch in {\it UVW2}, and only shows a very small rise in {\it UVW1}. 
In season two ($\sim$+230 to +460\,d), the {\it H} and {\it Ks} light curves appear to flatten, with {\it Ks} perhaps even rising slightly and {\it J} band behaving more similarly to the optical bands. At late times, from the third season onwards ($\gtrsim$+600\,d), a further flattening in the light curve becomes evident, with a slight decline of about 0.1 mag (100\,d)$^{-1}$ in the $V$ band. We note here that at these later times there may be a significant host contribution to the photometry, that we are unable to remove given the poor depth of the available templates. However, the flattening is also evident in the {\it Gaia} light curve which is expected to be largely unaffected by background contamination, and so we regard it as a real effect. The decline rate is much slower than that expected from the decay of radioactive $^{56}$Co (0.98 mag (100\,d)$^{-1}$, \citealt{Miller2010}), suggesting that circumstellar interaction remains the dominant mechanism powering the light curve.

\begin{figure}
\includegraphics[width=\columnwidth]{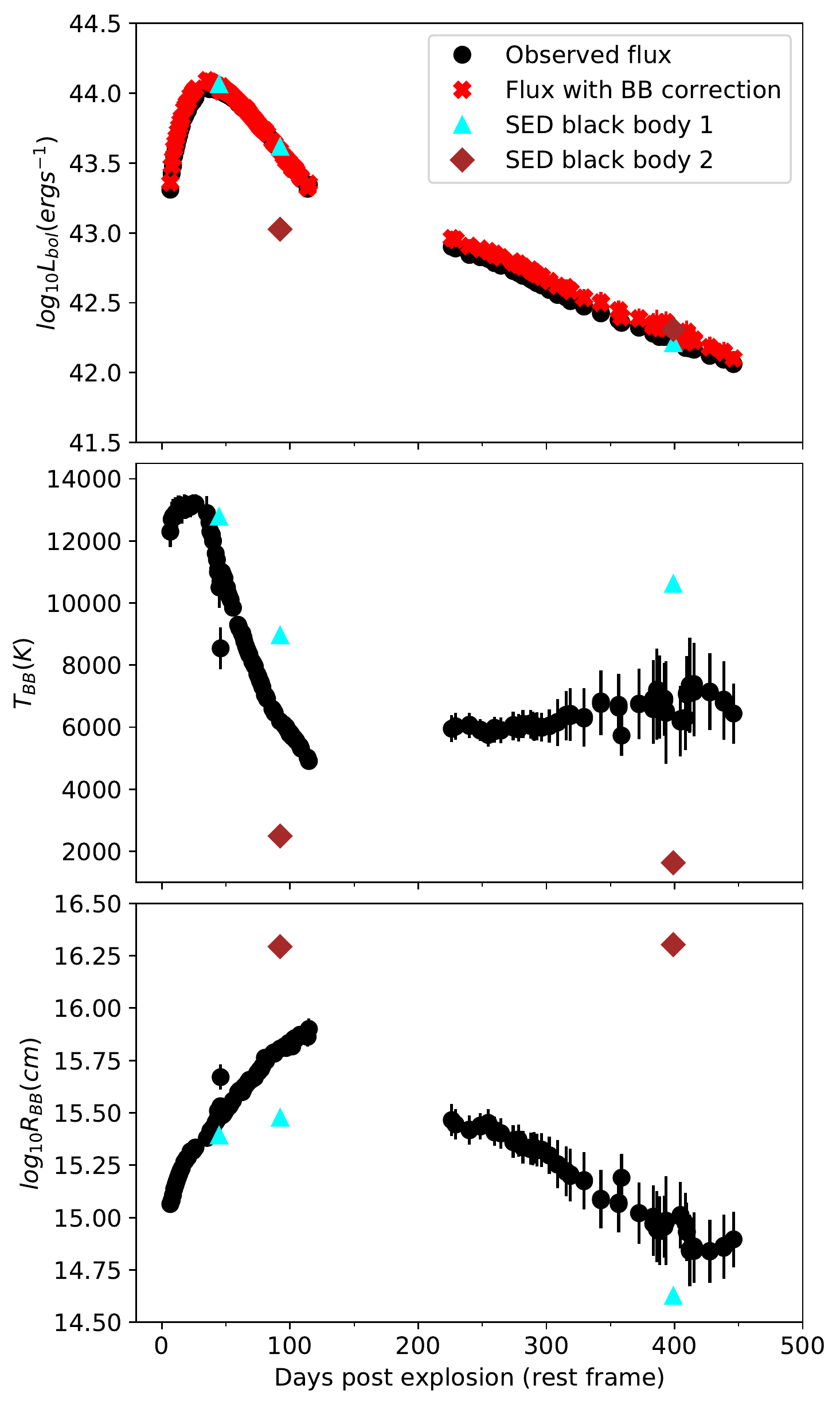}
\caption{Plot of black body fit parameters. Top panel: pseudo-bolometric light curve for SN~2017hcc, constructed from the $UVW2$, $UVM2$, $UVW1$, and $UBgcVroizJHKs$ bands. In construction of the pseudo-bolometric light curve, we removed points obtained from \texttt{Autophot} with S/N < 10. Middle panel: corresponding temperature evolution of the black body. Bottom panel: corresponding evolution of the black body radius. The cyan triangles and brown diamonds represent the black body fits shown in Fig. \ref{fig:SED_fit}.
}
\label{fig:bolometric}
\end{figure}
We compare SN~2017hcc to the slowly evolving SNe IIn SN~2010jl \citep{Stoll2011} and SN~2015da \cite{Tartaglia2020},
as well as to members of the class that are, at least initially, faster evolving, 
viz. SN~1998S \citep{Fassia2000} 
and SN~2005ip \citep{Stritzinger2012}.  
These objects were chosen due to their diversity and their well-sampled data sets, however there are many other examples of comparable objects. The details of the comparison objects can be seen in Table \ref{tab:compobjects}.
Other long-lasting SNe IIn with long rise times include PTF12glz, HSC16aayt (SN~2016jiu), and SN~2015da. 
PTF12glz had a rise time of about 50 days with photon diffusion through an aspherical CSM being offered as an explanation in that case \citep{Soumagnac2019}. HSC16aayt had an even slower rise, peaking at about $-$19.9 mag after over 100 days \citep{Moriya2019}. Similarly, SN~2015da had a rise time of 100 $\pm$ 5 days in the $R$ band, peaking at around $-$20.45 $\pm$ 0.55 mag, and was observable for at least five years \citep{Tartaglia2020}. 

\begin{figure}[ht!]
\includegraphics[width=0.95\linewidth]{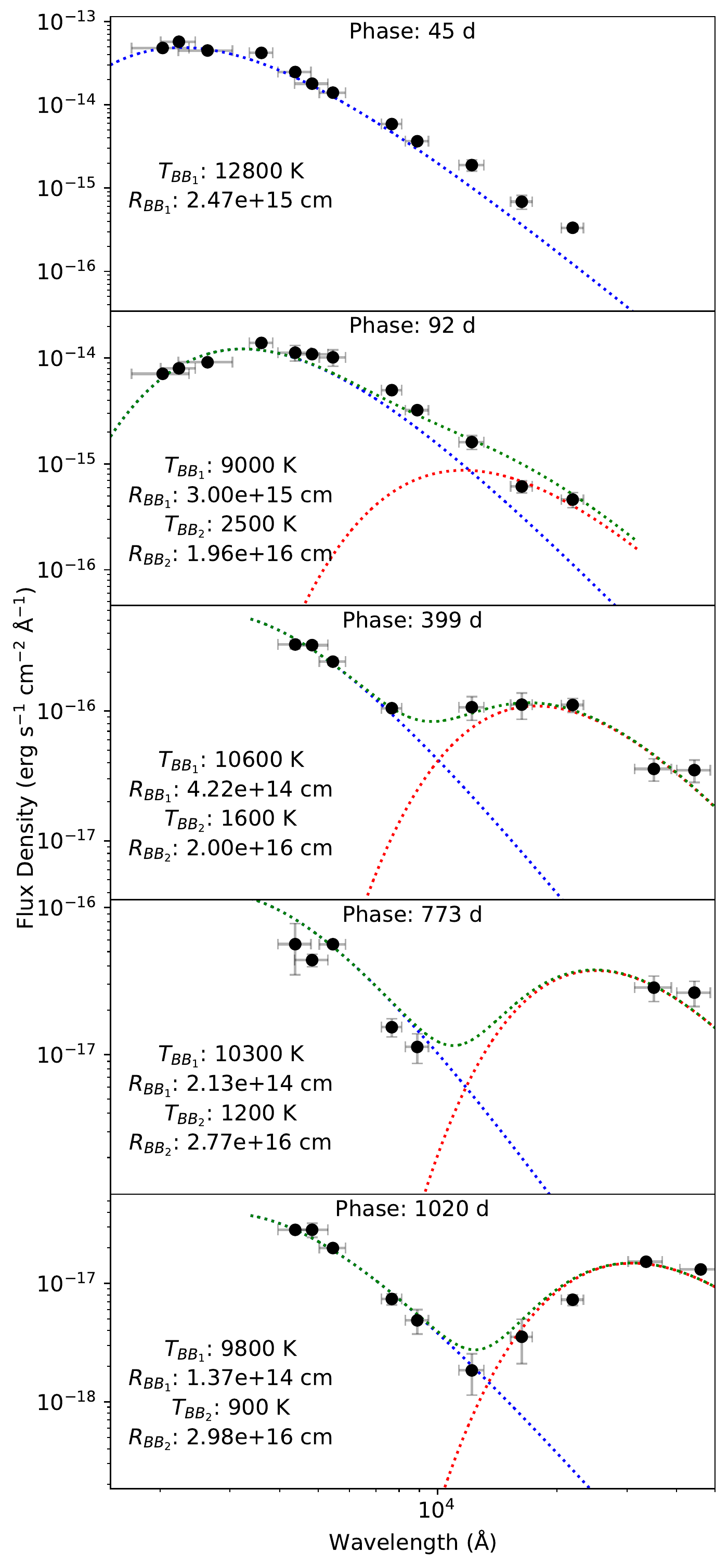}
  \caption{Black body fits to the spectral energy distribution. In the first two panels, representing early times, a single black body is sufficient to fit the SED, but in the final three panels, far later in the evolution of the SN, two black bodies are necessary. The $r$ band data points were not included due to the influence of the strong  H$\alpha$ emission in this range. In each panel all points are from epochs within ten days of the listed phase, after interpolation. It should be noted that the second black body +92\,d SED fit had an upper limit of 2500\,K on the possible model temperatures and the model determined by the MCMC fitting is very close to this value. The horizontal bars on each point represent the bandwidths of the filters. The temperatures and radii of the black body fits used are given in each panel. This has been corrected for foreground extinction.}
  \label{fig:SED_fit}
\end{figure}
The peak $r$ band magnitudes of our SN IIn comparison sample span the space from $-$18 to $-$21 mag and the light curves last a few hundred to over a thousand days (Fig. \ref{fig:lightcurve_comp}). A larger sample, such as that considered in \citealt{Nyholm2020}, shows that SN~2017hcc and SN~2010jl are exceptionally bright and long lived compared to more typical SNe IIn.

Turning to the early colour evolution (Fig. \ref{fig:col_curves}), we see a reasonably steady transition towards the red for SN~2017hcc in the $B-V$, $g-r$, $r-i$, and $V-K_\mathrm{s}$ colours over the first observing season (the first $\sim$150 days), consistent with the spectral evolution (Sect. \ref{fig:spectra}). The $B-V$ evolution of SN~2005ip is not substantial, but SN~1998S behaves similarly to SN~2017hcc with its swift movement redwards. The $B-V$ evolution of SN~2015da is rather similar to that of SN~2017hcc, though it remains redder throughout. In $g-r$ SN~2005ip and SN~2015da, like SN~2017hcc, are seen moving towards the red, though the shapes of the early evolution of SN~2005ip and SN~2015da differ from that of SN~2017hcc, with SN~2005ip quickly moving to a plateau before rising again and the evolution of SN~2015da being more gradual. The early evolution of SN~1998S in $V-R$ is very similar to that of SN~2017hcc in $g-r$. The early colour evolution of SN~2010jl contrasts with that of SN~2017hcc in that it steadily moves bluewards in $r-i$. Like SN~2017hcc in $r-i$, the SN~2015da and SN~1998S $R-I$ colours drift towards the red at early times before turning back towards the blue. SN~2017hcc has a strong shift towards the red in $V-K_\mathrm{s}$ in season one, but without any corresponding blueshifted spectral line evolution, which would often be seen in the case of dust formation in the ejecta. The $V-K_\mathrm{s}$ colour of SN~1998S has a similar redwards evolution to SN~2017hcc during the period corresponding to season one of the SN~2017hcc observations, but SN~2010jl is fairly flat at this time, as is SN~2015da. At early times SN~2015da is redder than SN~2017hcc in all colours and appears to evolve more slowly, though also in a redwards direction.

The colour curves in season two ($\sim$+230\,d to +460\,d) are markedly different from those in the first season.
Looking at the $B-V$ and $g-r$ colours (see Fig. \ref{fig:col_curves}), we see SN~2017hcc trending back towards the blue, though the scatter is large in the case of the $B-V$ colour curve. SN~2015da has a similar trend towards the blue in $B-V$ during an equivalent period, despite its consistently redder colour, but in $g-r$ it is fairly stable at this time. The overall bluewards evolution of SN~2017hcc does not appear to be caused by an increase in photospheric temperature (see Fig. \ref{fig:spectra}, and discussion of the bolometric light curve in Sect. \ref{sect:bol}).
Rather, the colour evolution at this phase is likely due to changes in line emission which is affecting broadband photometry.
In $r-i$ it appears that SN~2017hcc remains at a roughly constant colour during the second season, whilst SN~2010jl and SN~2015da evolve towards the blue at equivalent times.
Both SN~2010jl and SN~2015da show redwards evolution in $V-K_\mathrm{s}$ during the period corresponding to the second season of SN~2017hcc.

SN~2017hcc and SN~2015da appear to plateau in $B-V$ in season three (from $\sim$+600\,d to $\sim$+800\,d), whilst SN~2010jl becomes bluer. In $g-r$ SN~2017hcc shows a very slight trend towards the red with SN~2015da instead moving towards the blue, though it nevertheless remains far redder than SN~2017hcc. SN~2017hcc moves further redwards at this time in $r-i$, as does SN~2015da, whilst SN~2010jl instead continues on a trend of decreasing $r-i$ colour. 
SN~2017hcc is also significantly further to the red in $V-K_\mathrm{s}$ in the third season than in the second, though
later ($\sim$+1000\,d) we see that there has been movement back towards the blue. The redwards evolution of SN~2017hcc, initially present in all bands, is only maintained in the colours containing redder bands as time goes on, reflecting the decreasing temperature of the continuum and the increasing prominence of the infrared black body component (Fig. \ref{fig:SED_fit}).

\subsection{Bolometric light curve}
\label{sect:bol}

We constructed both pseudo-bolometric and bolometric light curves (Fig. \ref{fig:bolometric}) using {\sc Superbol}  \citep{Nicholl2018}, and including the following bands: {\it UVW2, UVM2, UVW1 and UBgcVroizJHKs}. We extrapolated and interpolated missing light curve data for a given filter assuming a constant colour relation with the nearest epochs; however, in the case of the {\it Swift} filters from {\it U} to {\it UVW2} we fitted a second order polynomial which we used for the later evolution in season one (the first $\sim$150 days), whilst we assumed that the colour was constant in the period before the first observations in those bands. We did not consider any contribution from the {\it U} to {\it UVW2} bands in later seasons (which were calculated separately), given the considerable drop-off in season one. After the initial peak there is a steady decline in the pseudo-bolometric light curve, as one would expect from the optical light curve. However, the temperature evolution panel in Fig. \ref{fig:bolometric} suggests an increase in temperature in the second season that begins a little after 200 days post explosion, and which is matched by a faster decrease in the radius. Whilst this could potentially be explained by the CSM interaction reheating the ejecta, this change does not appear to be reflected in the colour evolution, and it can be seen in Fig. \ref{fig:SED_fit} that a second black body becomes necessary to fit the spectral energy distribution (SED) at later times. It should be noted that the temperature of the first black body increases between +92\,d and +399\,d. This can be explained by a forest of Fe~{\sc II} lines, typical in SNe IIn, which affects the broadband photometry.

We examined the SED evolution of SN~2017hcc over time by fitting it with black bodies (Fig. \ref{fig:SED_fit}).
At early times a single black body is sufficient to fit the SED. However, as time goes on, a clear IR excess becomes evident and a second black body is required to fit this component. We did not include the $r$ band data in the SED, because the presence of the strong H$\alpha$ emission within this wavelength range artificially boosts the flux, preventing the description of the SED with a black body. SN~2005ip also displayed a growing IR excess with time, being fitted with two black bodies in \cite{Stritzinger2012} with the cooler black body becoming dominant by +100\,d. SN~2015da was another SN IIn showing such an excess; it required a second black body from +443\,d in order to reproduce the cooler component of the SED \citep{Tartaglia2020}. 

In order to explore the possible range of parameters that could generate the observed luminosity, we used the Python-based \texttt{TigerFit}\footnote{\url{https://github.com/manolis07gr/TigerFit}} code to fit models with a constant-density CSM shell to the bolometric light curve of SN~2017hcc. The \texttt{TigerFit} code makes use of semi-analytical light curve models described in 
\cite{Chatzopoulos2012, Chatzopoulos2013}, \cite{Kasen2010} and \cite{Dexter2013}, based on the approaches outlined in \cite{Arnett1980, Arnett1982}. However, despite trialling models with an extremely broad range of parameters, we were unable to simultaneously describe the peak, early evolution and late evolution in a manner consistent with observations, suggesting more sophisticated modelling is required.

\subsection{Infrared excess}
The development of an IR excess becomes evident from the $V-K_\mathrm{s}$ colour evolution soon after the peak (Fig. \ref{fig:col_curves}). We see that an IR excess is clearly apparent in the optical – IR SED by +92\,d from explosion. 
At this point we find it necessary to include a second black body component when fitting the SED (Fig. \ref{fig:SED_fit}) and we estimate that the IR component contributes roughly 13\% of the total luminosity obtained by summing the two black body components. By +399\,d the IR component is already very significant, contributing over 50\% of the total luminosity and this increases to over 70\% by +774\,d. We discuss this further in Section \ref{sect:disc_conc}. There will naturally be a time delay between the arrival of optical photons which are emitted directly towards the Earth and those which are absorbed by dust and then re-emitted in the IR \citep{Graham1986}, however we do not expect the effect of this to be very significant given the very long time scale of our light curve, so we do not account for it.

\section{Spectral evolution}
\label{sect:spec_evo}
\subsection{Optical spectra}
\label{subsec:optical_spectra}
\begin{figure*}
\includegraphics[width=1\linewidth]{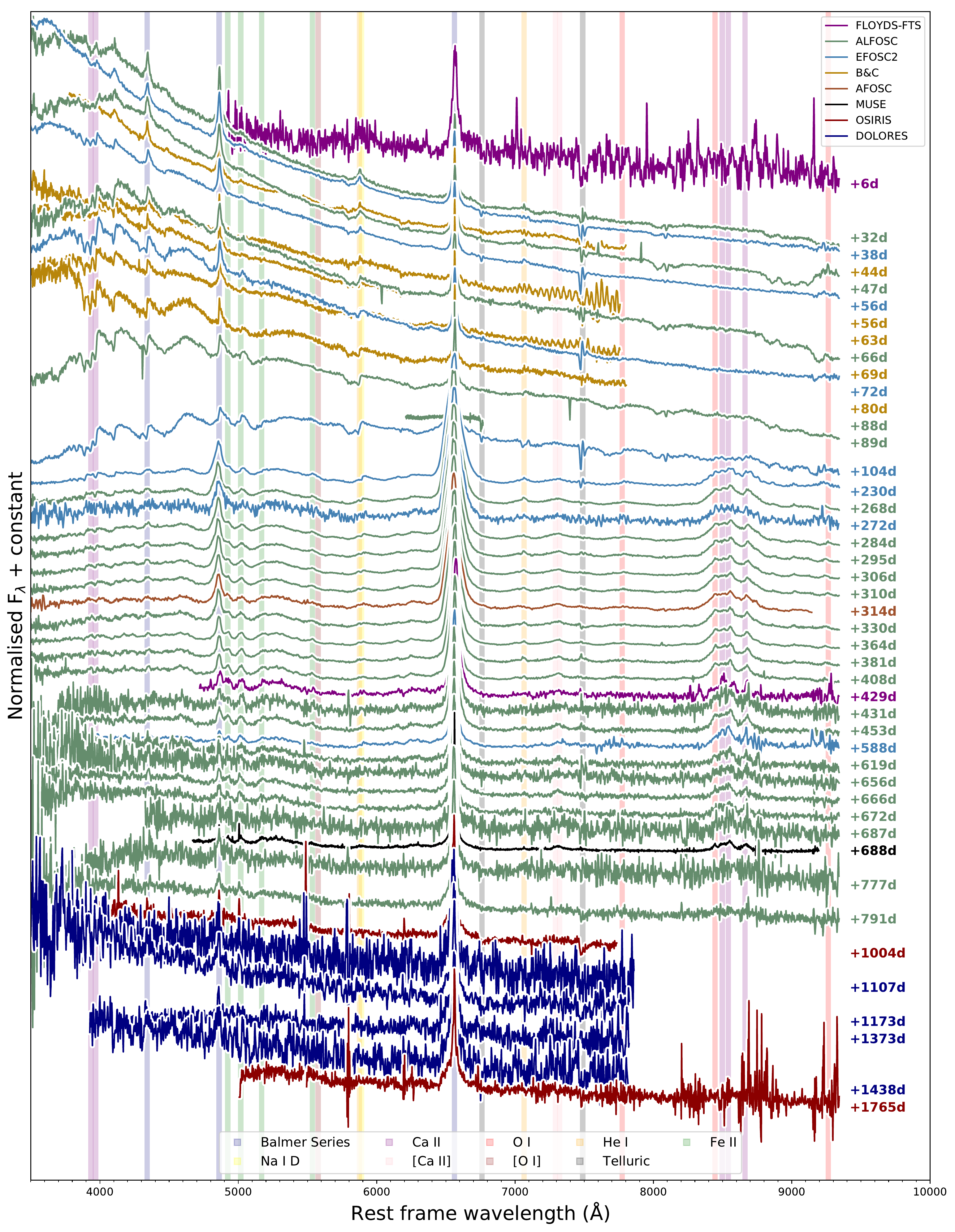}
\caption{Full sequence of optical spectra of SN~2017hcc, plotted in the SN rest frame. The wavelengths of some of the stronger SN lines are marked with coloured bands. The flux values for each spectrum have been normalised against the peak of H$\alpha$.}
\label{fig:spectra}
\end{figure*}

The optical spectra of SN~2017hcc span a period of 1759 days (4.8 years): from six to 1765\,d post-explosion (Fig. \ref{fig:spectra}). The overall evolution is generally slow, in line with the photometric evolution.

\begin{figure}
\includegraphics[width=\columnwidth]{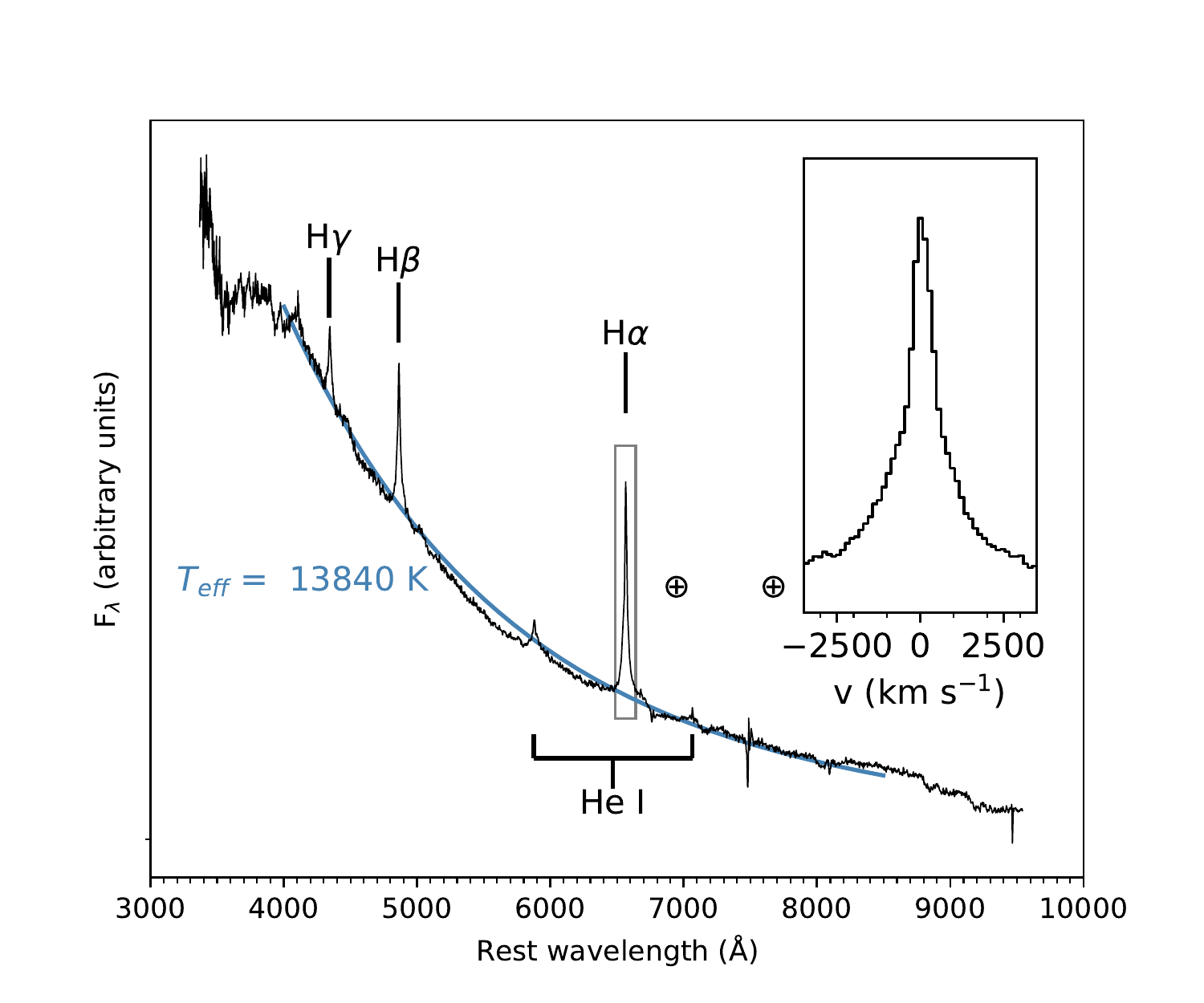}
  \caption{Early (+32\,d; 2017 Nov 1) spectrum fitted with a black body. Identified emission lines are labelled, and an inset shows a zoom in on the H$\alpha$ line in velocity space.}
  \label{fig:spec_early_ID}
\end{figure}

The spectra are initially characterised by a strong blue continuum and are dominated by strong Balmer emission, as is typical for a SN IIn. This can be seen clearly in our spectrum at +32\,d (Fig. \ref{fig:spec_early_ID}), where the H$\alpha$, H$\beta$, and H$\gamma$ lines are visible. H$\alpha$ is well fitted by a single Lorentzian profile with a FWHM of 1170 $\pm$ 20~\kms.
In addition, we see He~{\sc i} at $\lambda$5876 and $\lambda$7065 with a weak emission component and a P~Cygni minimum at 4000~\kms.

\begin{figure}
\includegraphics[width=\columnwidth]{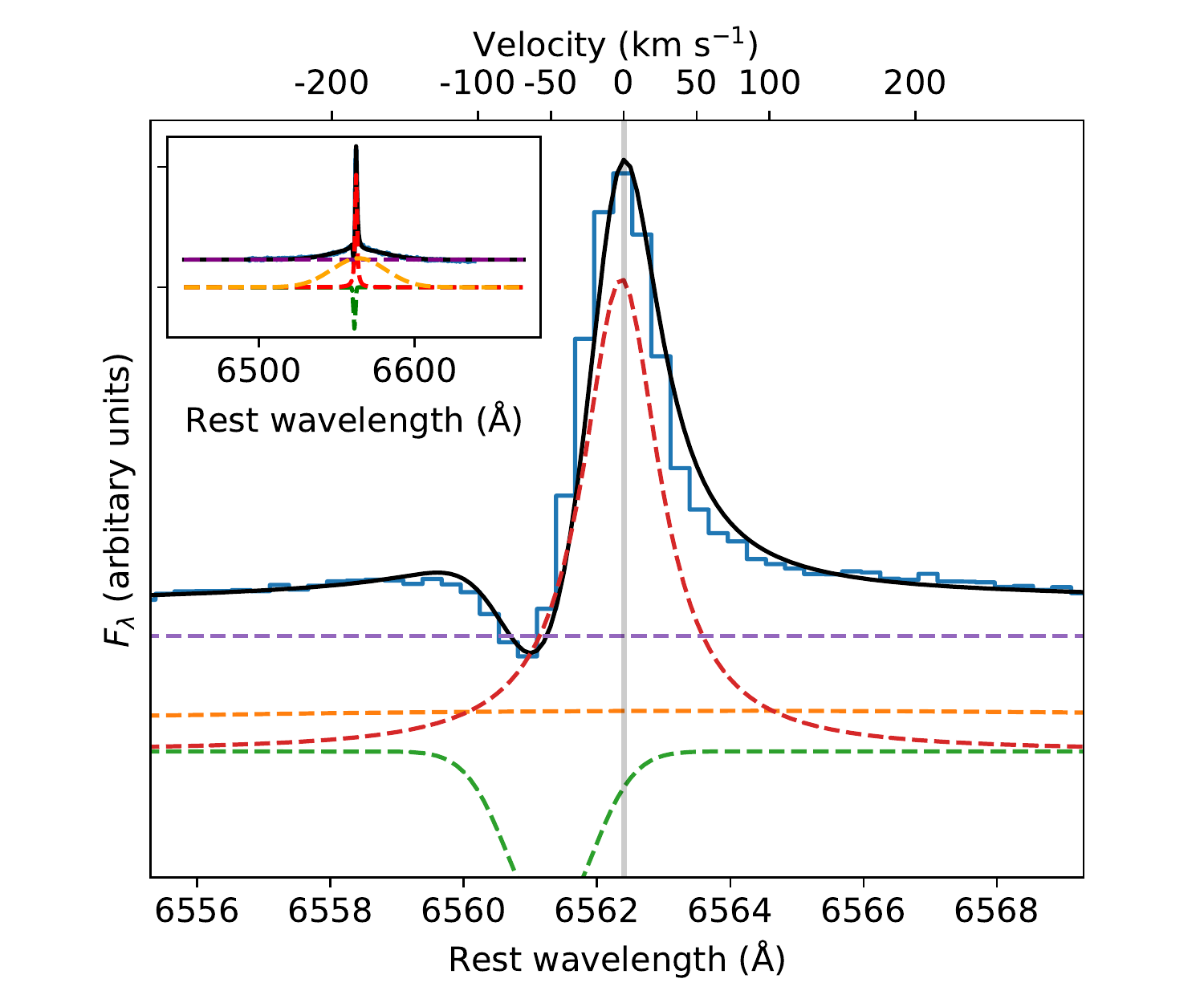}
  \caption{Fitting the +88\,d H$\alpha$ profile taken with the medium-resolution NOT Gr\#17 spectrum (blue). Dashed lines show the individual components: the orange and green lines are Gaussian fits to the broad component and absorption trough, respectively; the red is a Lorentzian profile fit to the narrow emission component, and the purple line represents the underlying continuum. The solid black line shows the composite fit. The inset shows a wider wavelength range revealing the broad wings in H$\alpha$. The orange component in the inset has been multiplied by a factor of 3 to more clearly demonstrate its shape. The vertical line marks zero velocity.
  }
  \label{fig:spec_midres}
\end{figure}

Fitting a black body to the +32\,d spectrum we find an effective temperature of 13.8 $\pm$ 0.1\,kK. However, this cools rapidly over the following week, dropping to 11.1 $\pm$ 0.1\,kK at +38\,d (we see a consistent trend in the evolution of the pseudo-bolometric light curve, see Sect. \ref{sect:bol}). 
The He~{\sc i} $\lambda$5876 emission also weakens, relatively, over the first two months. 
The Fe~{\sc II} $\lambda$5018 line is apparent from at least the +32\,d spectrum in season one.
In Fig. \ref{fig:spec_midres} we show a medium-resolution ({\it R}$\sim$5000) NOT+ALFOSC Gr\#17 spectrum taken at +88\,d covering the H$\alpha$ region. The spectrum clearly shows a narrow P~Cygni absorption, indicative of absorption by slow moving optically thick CSM along the line of sight.
In order to measure velocities of this feature, we fitted it with a four component model to describe the narrow and broad emission and the narrow absorption, as well as the continuum. The components were chosen purely due to their ability to reproduce the profiles seen in the data. 
These components are shown separately in Fig. \ref{fig:spec_midres}, and the composite model does an excellent job of reproducing the observed H$\alpha$ profile, though there are a lot of parameters. 
The narrow Lorentzian has a FWHM of $\sim$63~\kms, 
whilst the P~Cygni absorption has a minimum velocity of $\sim -$51~{\kms} with respect to the peak of the Lorentzian which we take as the pre-SN wind velocity of the progenitor. The FWHM of the broad Gaussian we fitted to the emission is $\sim$795~\kms. The aforementioned values have not been corrected for instrumental resolution and it is likely that the narrow emission feature that we fit with the narrow Lorentzian is not truly resolved.

At approximately +120\,d, SN~2017hcc disappeared behind the Sun. When it re-emerged, at $\sim$+230\,d, the H{$\alpha$} emission had become broader, with a velocity of 1690 $\pm$ 20~{\kms} at $\sim$+270\,d and the centre of the emission was clearly blueshifted with respect to the rest wavelength. The [Ca~{\sc II}] $\lambda\lambda$7291,7323 lines became visible in the +230\,d spectrum, with a velocity broadly similar to that of H{$\alpha$}.
At some point between 104 and 230 days post-explosion the Ca~{\sc ii} $\lambda\lambda\lambda$8498,8542,8662 triplet appears. 
The triplet grows in relative strength over the course of the next 200 days. The He~{\sc i} $\lambda$7065 line appears to have grown in relative prominence by +230\,d (the He~{\sc i} $\lambda$5876 line also appears to strengthen, but this is affected by blending with Na~I~D). 
It continues to be present throughout season two, though appears to weaken in relative prominence over the course of the season. After about +400\,d we see a pseudo-continuum beginning to form in the blue. Many emission lines, primarily iron, combine in raising the flux bluewards of about 5000 \AA. Similar behaviour has been seen in other SNe IIn (e.g. SN~1995N, \citealt{Fransson2002}; SN~2005ip, \citealt{Stritzinger2012}). We do not see a clear bluewards evolution in the $B-V$ colours at late times, however there are relatively large uncertainties on our photometric measurements at this phase.
In the +230\,d spectrum the Fe~{\sc II} $\lambda$4924 and $\lambda$5018 lines are evident, with the $\lambda$5169 component of the multiplet becoming evident only later, after around +300\,d. 

\begin{figure}
\includegraphics[width=\columnwidth]{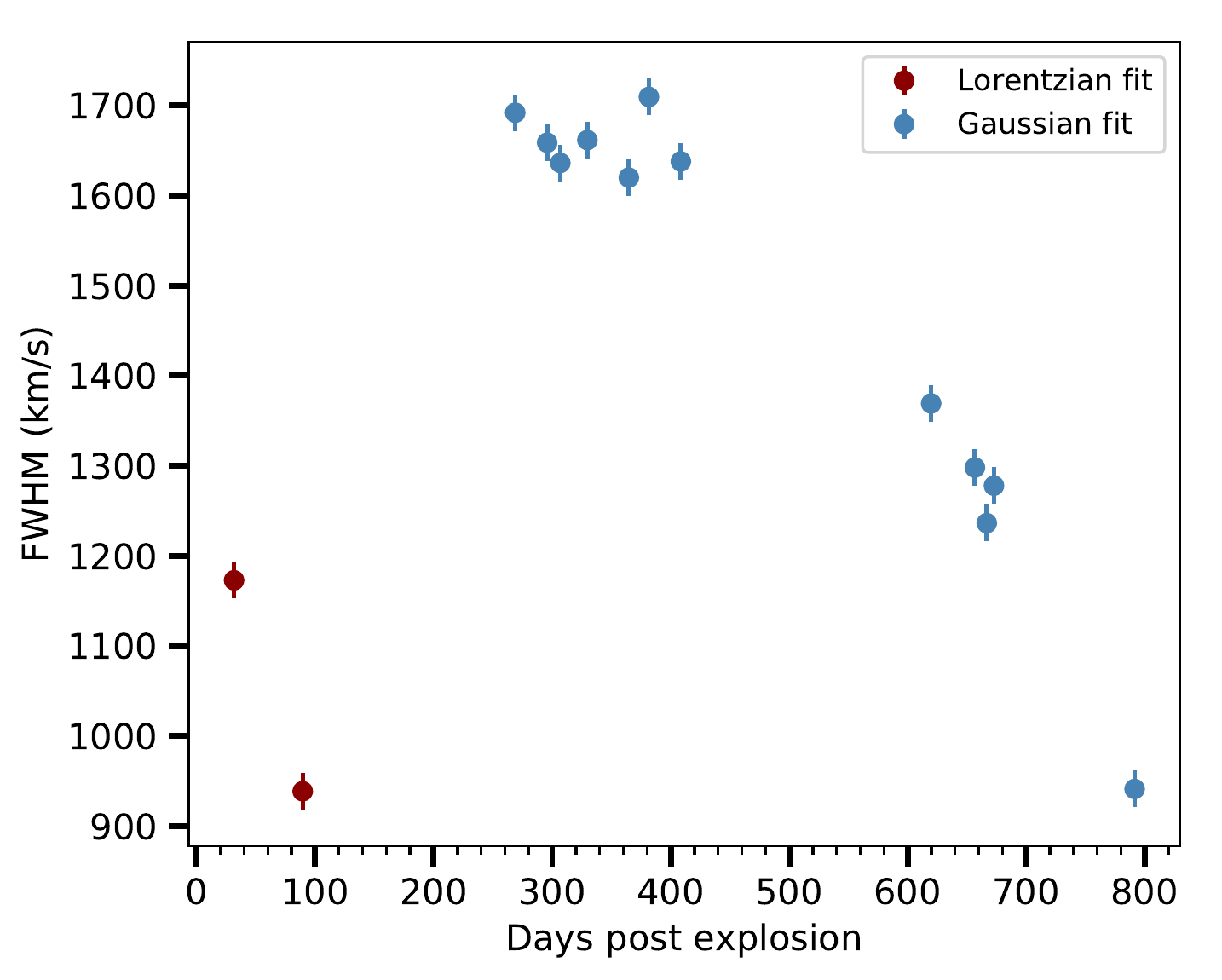}
  \caption{Evolution of the H$\alpha$ FWHM in the NOT+ALFOSC Gr\#4 1.0$''$ spectra over time. For each spectrum in the second season (after around +230\,d) we fitted two Gaussian profiles to H$\alpha$, but those corresponding to the broad base are not shown here. As alluded to above, all spectra were taken with an identical setup. The values have been corrected for instrumental resolution. The errors on each point represent a characteristic error, determined by repeated sampling of the skyline and H$\alpha$ FWHM of the +306\,d spectrum.}
  \label{fig:spec_fwhm_evol}
\end{figure}

\begin{figure}
\includegraphics[width=\linewidth]{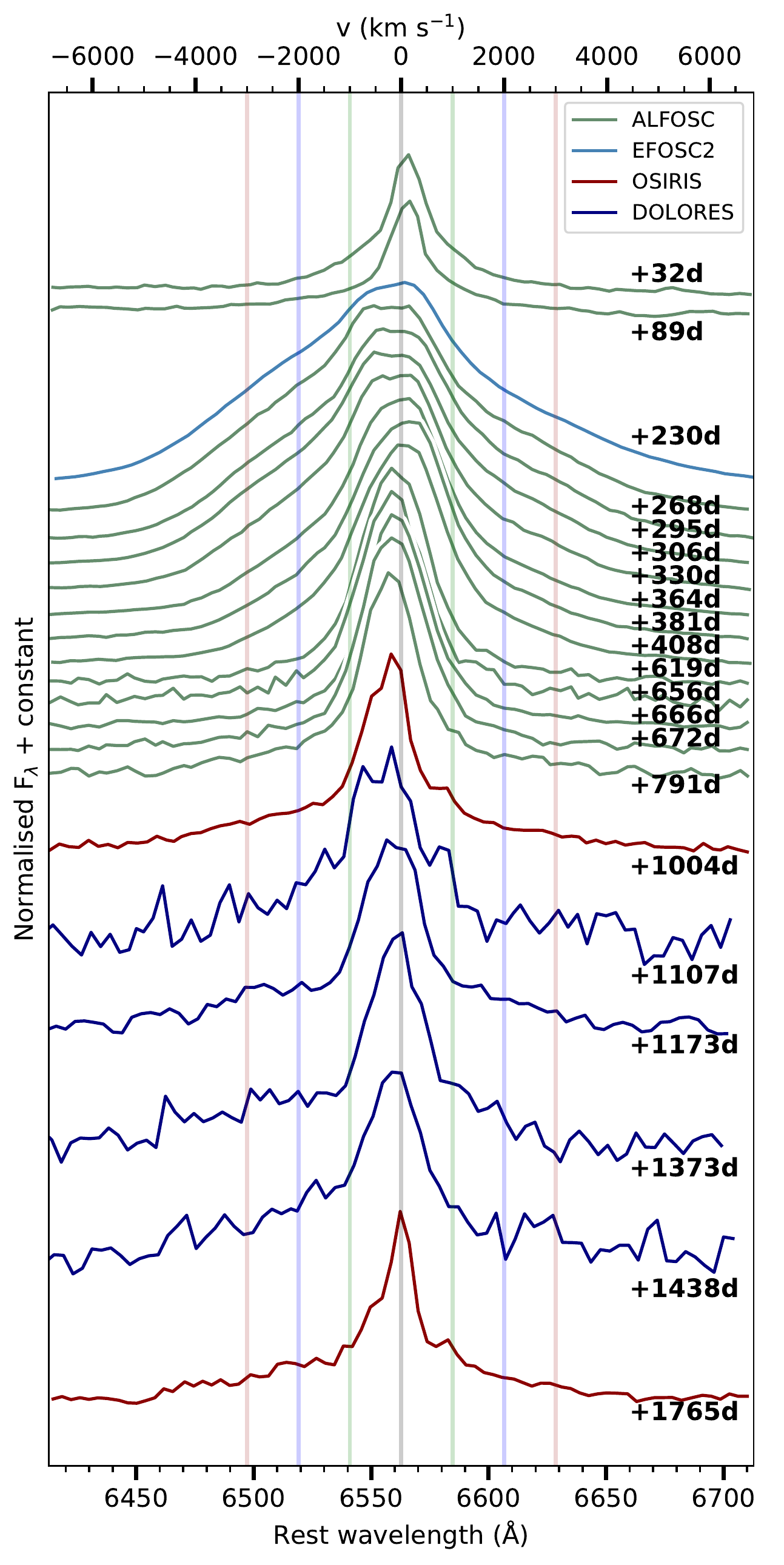}
\caption{
Selection of spectra showing the evolution of the H$\alpha$ profile. 
The ALFOSC spectra (green) were taken with an identical set up. The other spectra were taken with different instruments, but all have been rebinned to a common resolution (Table \ref{tab:spec}). The central vertical line is at the rest wavelength of H$\alpha$, whilst the vertical green, blue, and brown lines are 1000, 2000, and 3000~\kms respectively from the rest wavelength. The flux values for each spectrum have been normalised against the corresponding peak value of H$\alpha$. The onset of a blueshift is apparent from +230\,d onwards.}
\label{fig:halpha_velocity}
\end{figure}

The [Ca~{\sc ii}] $\lambda\lambda$7291,7323 lines are no longer visible in the +672\,d spectrum. The Ca~{\sc ii} $\lambda\lambda\lambda$8498,8542,8662 triplet weakens throughout the third season, being only barely evident in the +791\,d spectrum. 
The Ca~{\sc ii} triplet is noticeably weaker than its final season two appearance even in the first season three spectrum to include the relevant wavelength range. The He~{\sc i} $\lambda$5876 and $\lambda$7065  emission lines appear to fade by +791\,d. 
As H$\alpha$ is the strongest line that is visible throughout the evolution of SN~2017hcc, we use it to investigate the velocity evolution (see Fig. \ref{fig:spec_fwhm_evol}). For the purpose of consistency, we only used the 1.0" slit spectra from the ALFOSC instrument. The H$\alpha$ FWHM is measured to determine the velocity of the post-shock material, as such it is the intermediate-width component with which we concern ourselves. 
For each spectrum we measure the instrumental resolution from the FWHM of the [O~{\sc I}] $\lambda$5577.34 sky emission line. 
We fitted Lorentzian and/or Gaussian functions to H$\alpha$ and determined their FWHM values which were then corrected for instrumental resolution. 
The uncertainty in our measurements is dominated by the choice of continuum region, so we sampled the FWHM of the +306\,d spectrum eight times in order to choose a representative standard deviation, which we then applied to all epochs. The first two epochs are well matched by a single Lorentzian, indicative of more electron scattering at early times due to denser CSM. At later times we instead fitted two Gaussians, with one being used to account for the broad base seen in Fig. \ref{fig:halpha_velocity}; however, the use of this Gaussian was purely a functional choice and we ascribe no physical significance to it. 
We can see that the velocity in the first epoch is around 1170~\kms, whilst a much higher plateau of about 1650~\kms is seen in the second season. Later on, there is a gradual decline before reaching approximately 940~\kms by around +800\,d. 

In Fig. \ref{fig:halpha_velocity} we show the evolution of H$\alpha$ over time. A clear blueshift, also noted by \cite{Smith2020}, is apparent in our data from +230\,d, but this may be decreasing in our spectra by +1173\,d. The +230\,d H$\alpha$ profile itself appears to have a flat top that persists out to $\sim$+410\,d. The flat top and blueshift are also present in the H$\beta$ profile. We discuss this further in Section \ref{sect:disc_conc}.

\begin{figure}
\includegraphics[width=\columnwidth]{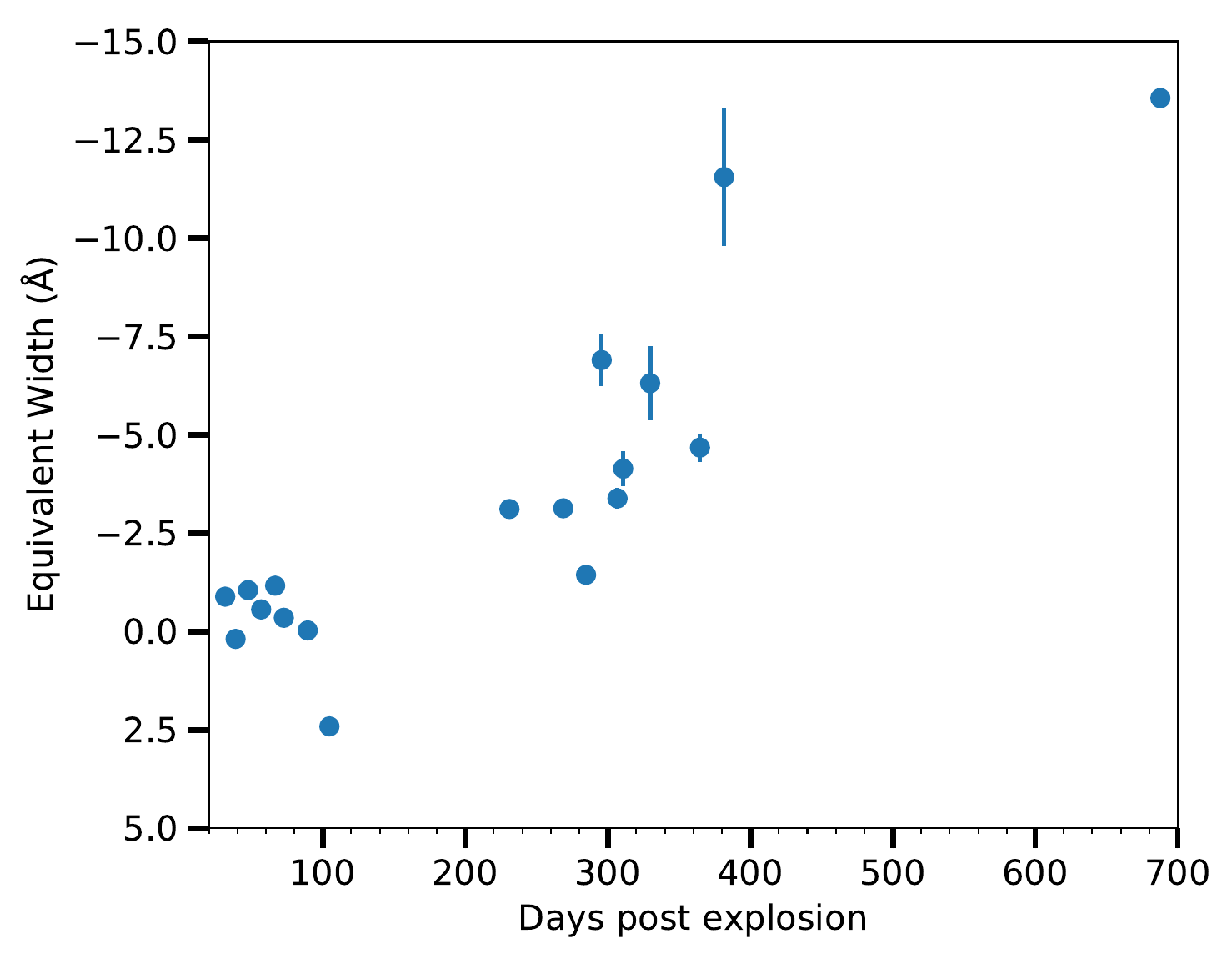}
  \caption{Evolution of the He~{\sc I} $\lambda$7065 line equivalent width over time. We only show those epochs for which the signal-to-noise ratio of the equivalent width measurement is greater than 5.} 
  \label{fig:spec_EW_HeI_evol}
\end{figure}

In Fig. \ref{fig:spec_EW_HeI_evol} we show the evolution of equivalent width of the He~{\sc I} $\lambda$7065 line. This region of the spectrum is relatively free of contamination from other lines throughout the evolution. Having chosen a region wide enough to encompass the line, we fit a continuum across it, and integrate the area underneath following \citet{Ebbets1995}. A clear increase in the absolute value of the equivalent width over time is visible. As this occurs even with falling temperatures we attribute it to greater non-thermal excitation of He~{\sc I} in CSM by increased CSM interaction.

\subsection{Optical spectra comparison}

\begin{figure}
\includegraphics[width=\linewidth]{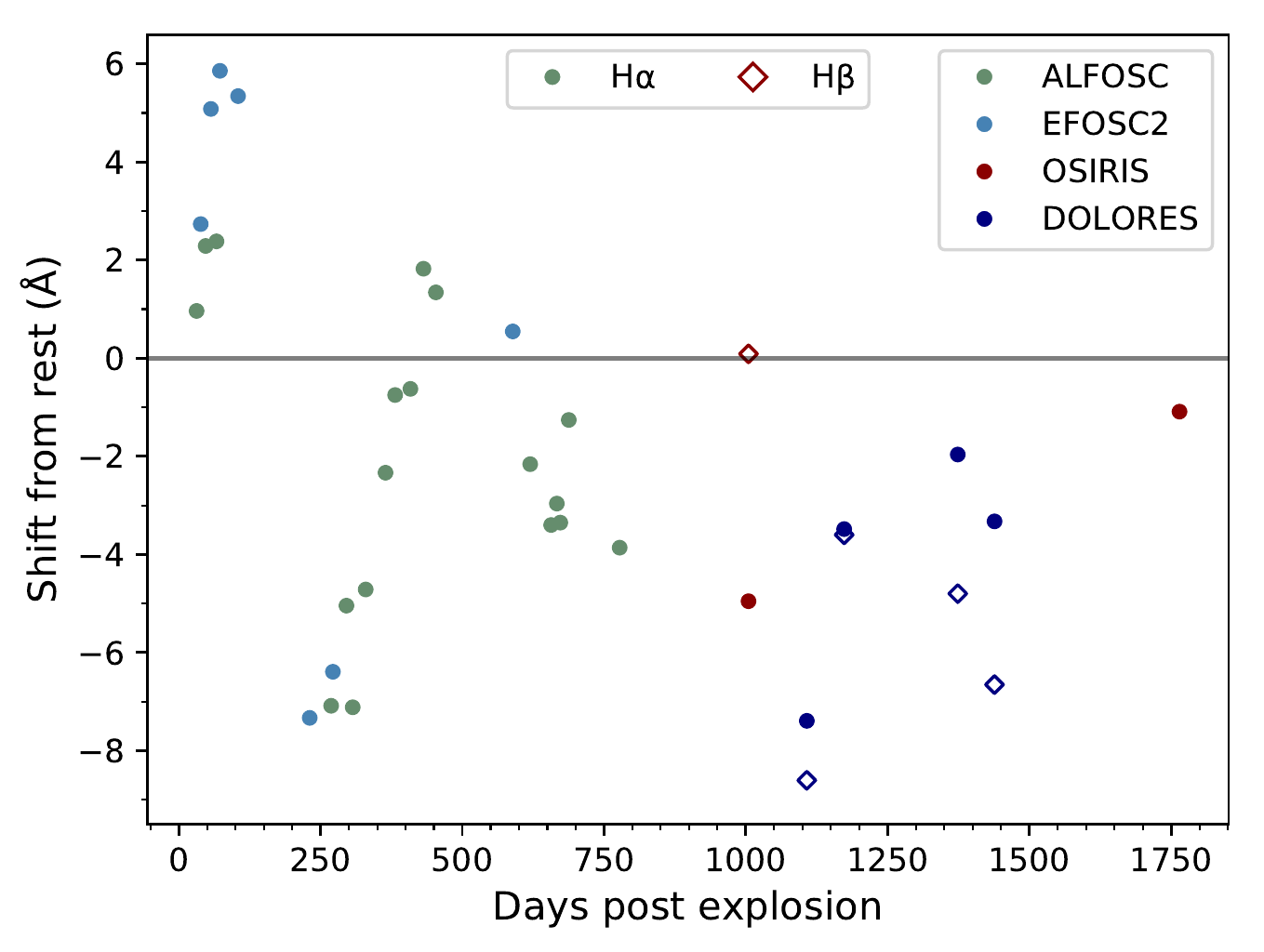}
\caption{
Selection of spectra showing the evolution of the H$\alpha$ line centre. The centres have been shifted slightly based on the position of the 6300.31 {\AA} skyline, except in the case of +32\,d and +330\,d ALFOSC spectra and the +272\,d EFOSC spectrum which have been shifted based on the position of the the 5577.34 {\AA} line as the 6300.31 {\AA} line was not detectable.}
\label{fig:halpha_centre}
\end{figure}

\begin{center}
\begin{figure*}
\includegraphics[width=1.96\columnwidth]{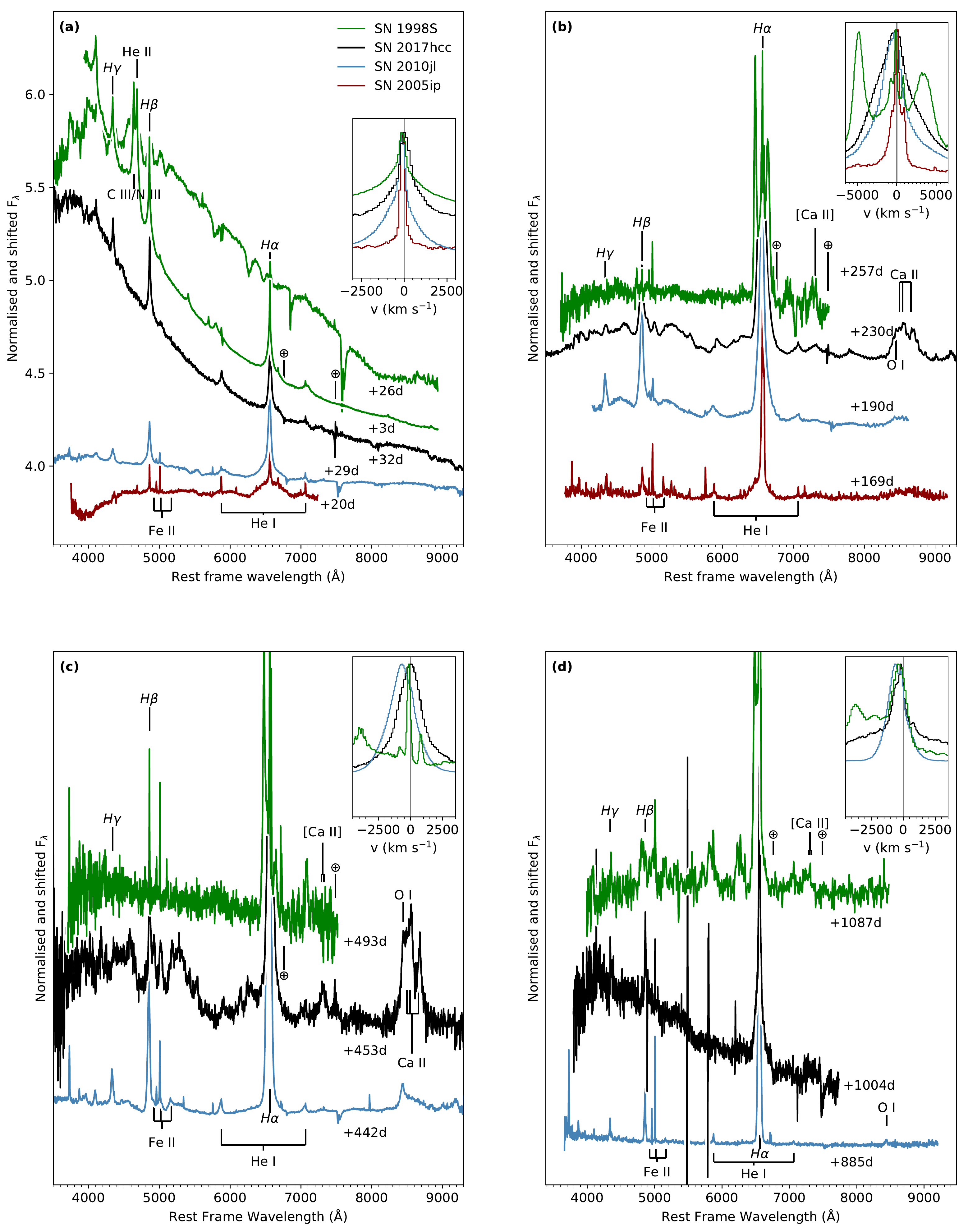}
  \caption{Spectral comparison of SN~2017hcc with SN~2010jl \citep{Zhang2012, Jencson16}, SN~2005ip \citep{Stritzinger2012} and SN~1998S \citep{Leonard2000, Fransson2005} at four selected epochs. The spectra have been corrected for redshift. The inset in each panel shows H$\alpha$ with the rest velocity marked with a vertical line to make the blueshift more obvious. The telluric lines marked are those corresponding to the SN~2017hcc spectra. Phases are given against discovery date, except in the case of SN~2017hcc, where the explosion epoch is used. The flux values for each spectrum have been normalised against the corresponding peak value of H$\alpha$.}
\label{fig:spectra_comp}
\end{figure*}
\end{center}

We now consider how SN~2017hcc compares to the three other well-observed type IIn SNe shown in Fig. \ref{fig:spectra_comp}. We see a strong similarity between the spectrum of SN~2017hcc taken at +32\,d and the spectrum of SN~1998S taken 3 days after explosion, whilst SN~1998S spectra at more similar epochs (e.g. +26\,d) are quite distinct. 
Both spectra have a noticeably blue continuum and their emission features are similar; however, the spectrum of SN~1998S shows high ionisation lines of C~{\sc iii} $\lambda$4648/N~{\sc iii} $\lambda$4640 and He~{\sc ii} $\lambda$4686, which are absent from the SN~2017hcc spectrum. These differences may be due to the early phase of the SN~1998S spectrum capturing photons coming from the shock breakout. This would explain the swift change in the appearance of the spectra of SN~1998S. 
SN~2010jl at a comparable epoch was much redder, though the overall emission features are similar to SN~2017hcc. SN~2005ip is also far redder than SN~2017hcc and has a broad component in the H$\alpha$ emission which is absent in SN~2017hcc. 
At these early times the light curve evolution of these objects is rather distinct, so differing spectral presentation is not surprising.
[Ca~{\sc ii}] $\lambda\lambda$7291,7323 is either very weak or completely absent in the +190\,d spectrum of SN~2010jl, but hints of this doublet can be seen in the +230\,d spectrum for SN~2017hcc and it is quite apparent by the +453\,d spectrum. 
The Ca~{\sc ii} $\lambda\lambda\lambda$8498,8542,8662 triplet is clearly present in the +230\,d and +453\,d spectra of SN~2017hcc. It is present in SN~2010jl in panel (c) too, but it is less prominent.

In the inset of panel (a) of Fig. \ref{fig:spectra_comp}, we see that the H$\alpha$ emission of SN~2010jl shows an excess in the blue. In the inset of panel (b), the H$\alpha$ profiles of both SN~2017hcc and SN~2010jl are blueshifted and are broader than that seen in SN~2005ip, which, along with SN~1998S, displays a complex multipeaked profile. A multipeaked profile, though not seen here, is visible in SN~2010jl at higher resolution \citep{Fransson2014, Jencson16}.

The H$\alpha$ of SN~2017hcc shows higher velocities than SN~2010jl and SN~2005ip. At late times, the H$\alpha$ velocities of SN~2010jl and SN~2017hcc are comparable, but SN~2010jl is clearly the more blueshifted of the two. The H$\alpha$ emission of SN~1998S continues to show a multipeaked structure with the central peak displaying a slight blueshift. The O~{\sc I} $\lambda$8446 line is clearly present in both SN~2017hcc and SN~2010jl at late times, but the SN~1998S spectrum does not extend to that wavelength, unfortunately.  

The presence of the Fe~{\sc II} $\lambda$5018 line can be seen in panel (a) in SN~2017hcc, SN~2010jl and SN~2005ip.
The Fe~{\sc II} $\lambda\lambda\lambda$4924,5018,5169 multiplet becomes visible in SN~2017hcc in panel (b) and is clearly present in panel (c) in both SN~2017hcc and SN~2010jl.

\begin{figure}
\includegraphics[width=\columnwidth]{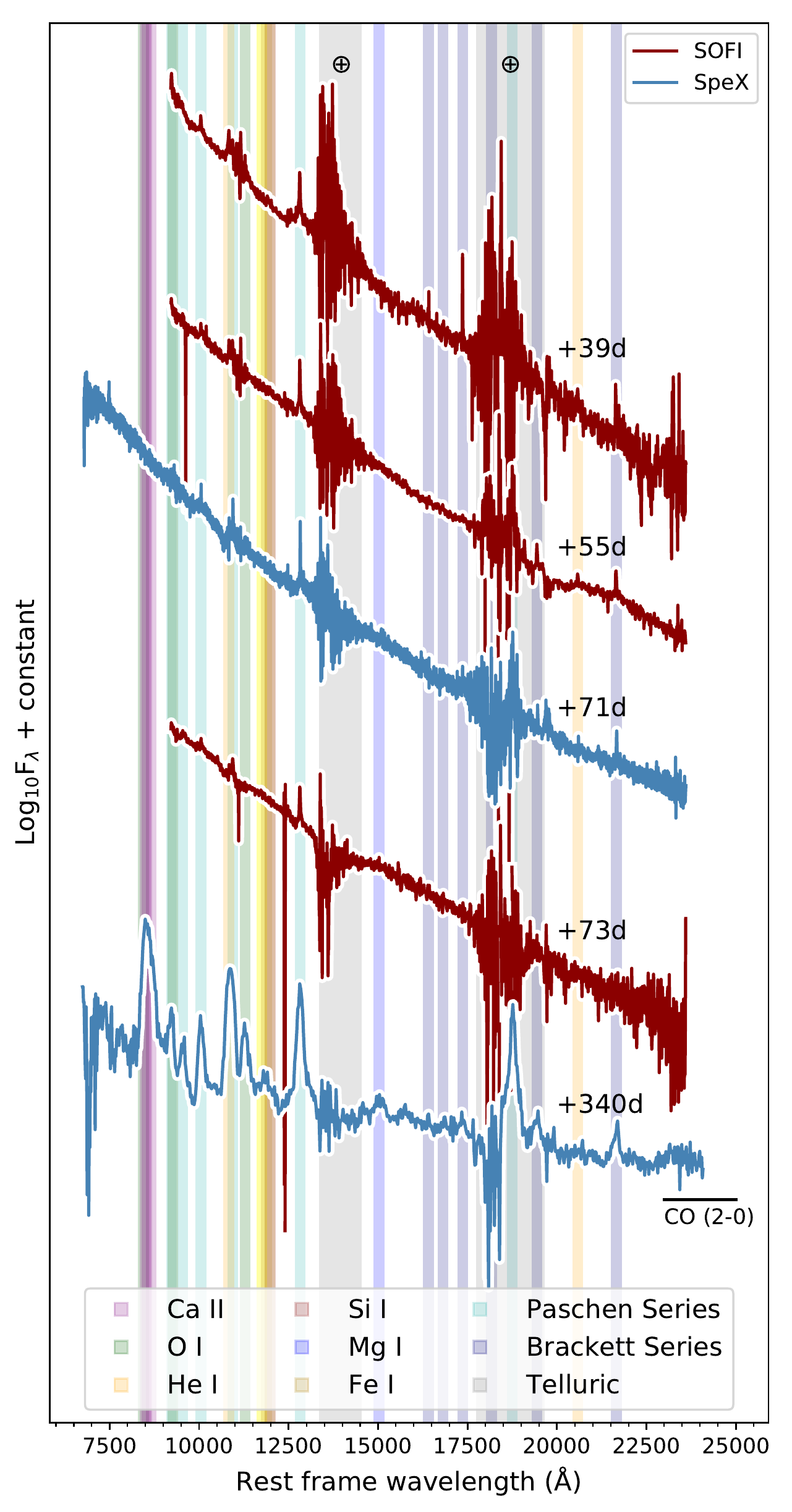}
  \caption{Complete sequence of NIR spectra for SN~2017hcc. The wavelengths of some of the stronger SN lines are marked with coloured bands and the two strong telluric absorption features are marked with pale grey columns.}

  \label{fig:NIR_spectra}
\end{figure}

\subsection{Infrared spectral evolution}
The NIR spectra (Fig. \ref{fig:NIR_spectra}) span a period of 301\,d, from  +39 to +340\,d, with the first four spectra covering a short period up to day 73 during which the spectral features change little: the spectra are rather blue, with the clearest lines from the Paschen and Brackett series, whilst some weaker features likely due to O are also apparent. The final spectrum, taken at +340\,d, shows significant evolution, and is markedly different from the earlier epochs. The Paschen and Brackett series now dominate the spectrum. Pa\,$\gamma$ appears to be blended with He~{\sc I} $\lambda$10830, and O~{\sc I} $\lambda$9264 and $\lambda$11287 are now also clearly discernible.
The latter is stronger than the either of the O~{\sc I} $\lambda$7774 and $\lambda$9264 features, possibly due to excitation by UV photons via Ly\,$\beta$. At this epoch, the first overtone of CO is also visible (Fig. \ref{fig:NIR_spectra}), and as with other core-collapse SNe, it probably appeared around +100\,d, although our data do not allow us to be more precise.

\begin{figure}
\includegraphics[width=\columnwidth]{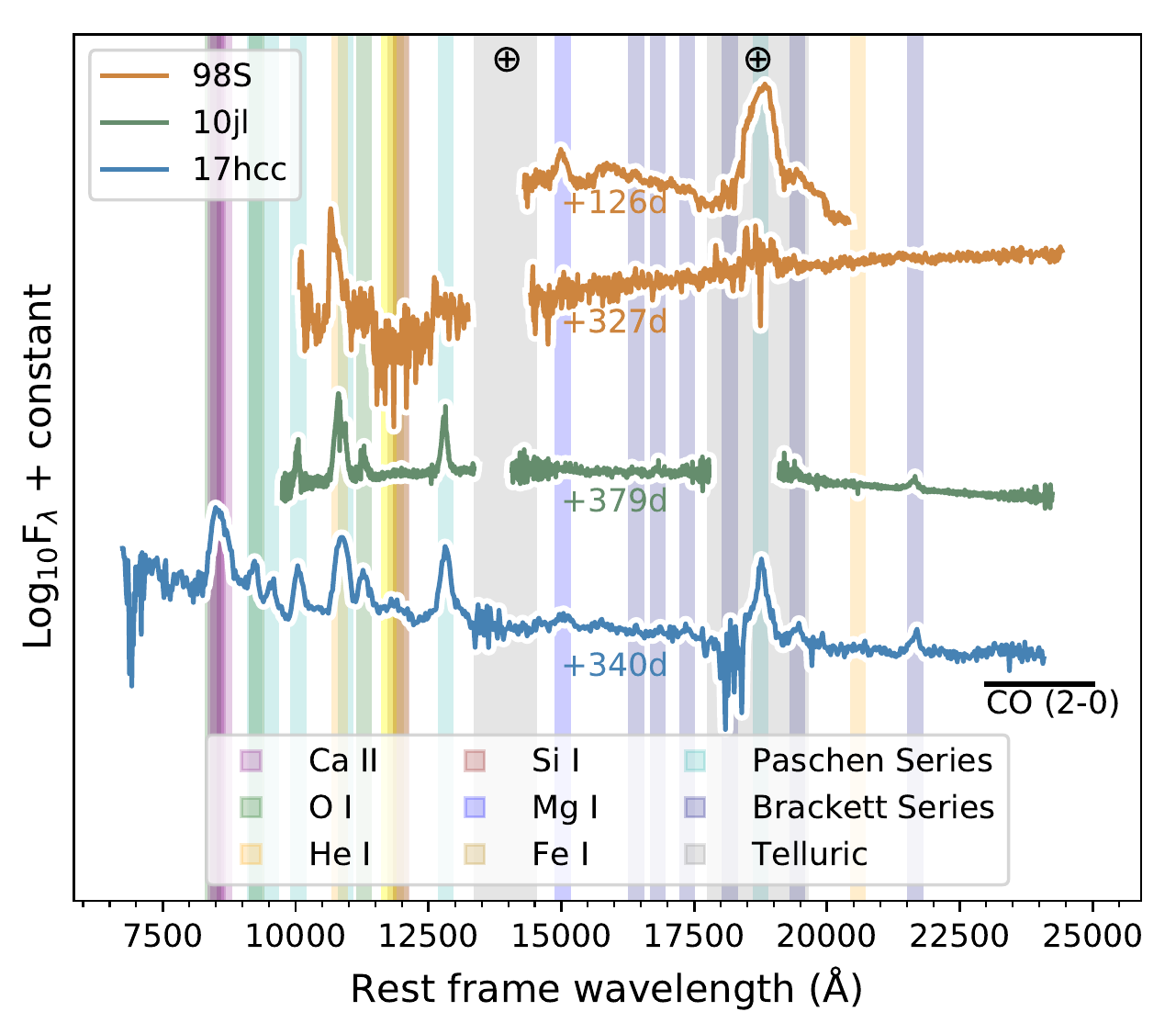}
  \caption{Comparison of the +340\,d spectrum of SN~2017hcc with late-time spectra of SN~1998S \citep{Fassia2001, Pozzo2004} and SN~2010jl \citep{Borish2015}. Phases are given against discovery date, except in the case of SN~2017hcc, where explosion epoch is used.}
  \label{fig:NIR_spec_comp}
\end{figure}

\begin{figure}
\includegraphics[width=\columnwidth]{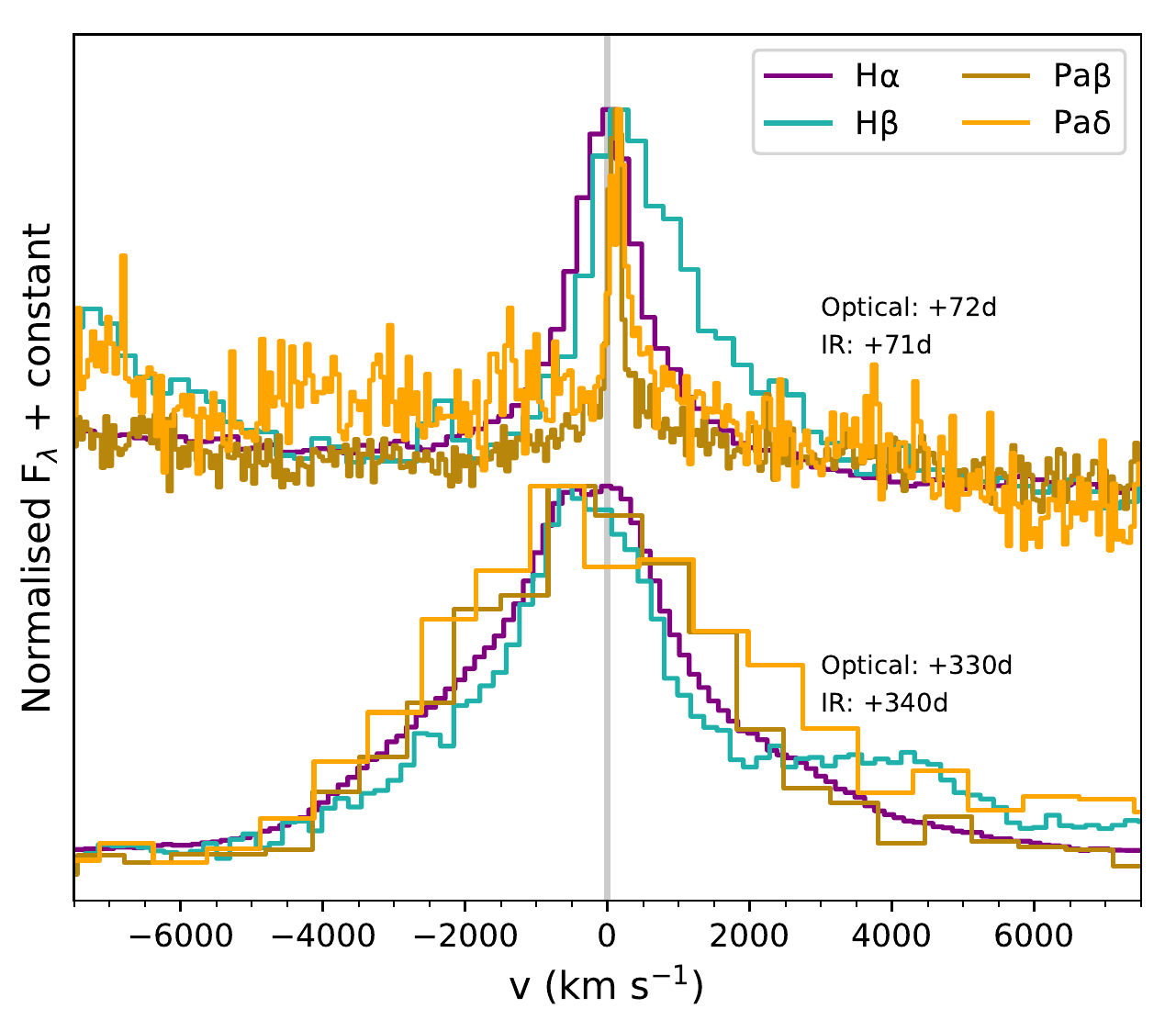}
  \caption{Comparison of the H$\alpha$, H$\beta$, Pa$\beta$, and Pa$\delta$ lines of SN~2017hcc at different epochs (+72\,d and +330\,d for the optical and +71\,d and +340\,d for the NIR). The Balmer and Paschen lines at comparable epochs have been overplotted. The lines have been normalised against their peaks. The vertical line marks zero velocity.}
\label{fig:Balmer_Paschen_overplot}
\end{figure}

In Fig. \ref{fig:NIR_spec_comp} we compare the +340\,d spectrum of SN~2017hcc with those of other type IIn SNe at comparable epochs. 
The +379\,d spectrum of SN~2010jl is quite similar to the SN~2017hcc spectrum in terms of the spectral features present. In the case of SN~1998S we show both the +126\,d and +327\,d spectra. The latter shows a clear red continuum, unlike SN~2017hcc, which instead shows a continuum more similar to that of SN~2010jl. The first CO overtone first appears in SN~1998S in a spectrum taken 109\,d after discovery \citep{Fassia2001}. There is no clear evidence for the presence of the first CO overtone in any of the NIR spectra of SN~2010jl seen in \cite{Borish2015}.

In Fig. \ref{fig:Balmer_Paschen_overplot} we compare the shapes of the H$\alpha$, H$\beta$, Pa$\beta$, and Pa$\delta$ profiles of SN~2017hcc. Those particular Paschen lines were chosen to avoid blending with other features. The difference in width of the optical and NIR lines at early times is likely to be due to differences in spectral resolution. At later times (+330\,d in the optical and +340\,d in the NIR) there is a clear blueshift present, as already noted for H$\alpha$ and H$\beta$ in \ref{subsec:optical_spectra}. The lines at late times are all in fairly good agreement with each other, though the Balmer lines do appear to display greater evidence of extinction, as would be expected from these bluer lines. The red shoulder visible in H$\beta$ is due to blending with Fe~{\sc II}.

\section{Metallicity}
\label{sect:metallicity}
We employed the O3N2 index, which uses the relative line strengths of [O III]/H$\beta$ and [N II]/H$\alpha$ to determine the metallicity of the host galaxy, as described in \cite{Pettini2004}. We measured these lines from the +688\,d MUSE spectrum. We extracted the spectrum with an aperture of radius 1.2 arc seconds centred on the core of the host. 
It should be noted that we determined the metallicity at the host galaxy centre due to signal-to-noise constraints, whilst it appears that the SN itself exploded in a spiral arm, which would be expected to be of lower metallicity \citep{Pilyugin2005}. We determined the metallicity in terms of oxygen abundance to be 12+log(O/H) = 8.49 $\pm$ 0.01 dex, that is to say approximately two-thirds solar \citep{Asplund2004}.

\section{Discussion}
\label{sect:disc_conc}

SN~2017hcc is a long-lived SN IIn that shows slow spectroscopic and photometric evolution as well as an IR excess, which becomes apparent soon after peak brightness. There are only a handful of type IIn SNe with five years of detailed observations.

The light curve shows a gradual rise to the peak (57 $\pm$ 2 days, ATLAS $o$ band) that is consistent with interaction with a massive and dense CSM, and can be explained by the diffusion timescale for photons through an optically thick CSM. The rise is followed by a monotonic decline, but after approximately two years the decline flattens out. There is no dramatic spectral change even at late times, and the emergence of nebular SN lines is not seen, suggesting that the optically thin inner ejecta never become visible, or that the ejecta density remains too high for forbidden lines to form.

SN~2017hcc shows an IR excess, which is apparent shortly following peak brightness, and can be seen in the $V-K_\mathrm{s}$ colour evolution (Fig. \ref{fig:col_curves}) as well as in the SED fits (Fig. \ref{fig:SED_fit}). Initially, a single black body is sufficient to match the SED, but a second, cooler, black body is needed to describe the SED after around 90\,d. 
Such behaviour is fairly typical, and has been observed in many type IIn SNe, such as SN~2005ip \citep{Stritzinger2012}, SN~2010jl \citep{Sarangi2018}, and SN~2015da \citep{Tartaglia2020}, to name but a few. 

Our analysis also included MIR photometry spanning 60 – 980 days and NIR photometry spanning 40 – 1170 days from explosion. At +110\,d the IR excess can be described by a T $\sim$ 1900\,K black body component consistent with the evaporation temperature for graphite dust grains. 
Our estimated radius for the IR black body at +110\,d of 2.3 $\pm$ 0.5 $\times$ 10$^{16}$ cm corresponds to a black body velocity of $\sim$24,000~{\kms}. The early onset of the IR excess together with the above values are suggestive of pre-existing dust in the CSM rather than condensation of new grains. 
Thus, these IR observations can be explained by absorption of the SN UV and optical photons and their re-radiation at IR wavelengths by the surrounding dust, that is to say an IR echo \citep[e.g.][]{Bode1979, Graham1983}.

Over the first $\sim$100 days, we can fit the intermediate-width H$\alpha$ emission with a single Lorentzian profile (Fig. \ref{fig:spec_fwhm_evol}), suggesting that for the first season the photosphere is located within the opaque CSM. After +200\,d the shocked ejecta start to become visible (when we see a FWHM of ~1650~\kms). Given that maximum velocities of over 5000~\kms\ are seen after +200\,d (Fig. \ref{fig:spec_fwhm_evol}), may suggest that we are seeing freely expanding ejecta as the CSM starts to fragment.

We also note that H$\alpha$ appears to have a flat top between $\sim-$700 and +300 \kms. Flat-topped or boxy profiles have been seen in a number of other SNe (e.g. SNe~1993J, 1995N \citealt{Matheson2000,Fransson2002}) where they have been accounted for as a sign of ejecta-CSM interaction, the flat top resulting from emission from a homologously expanding cool dense shell (CDS) formed between the forward and reverse shocks \citep[e.g.][]{Dessart22}. The appearance of the flat top is coincident with the increase in the equivalent width of the He~{\sc I} $\lambda$7065 line (Fig. \ref{fig:spec_EW_HeI_evol}) and the decrease in the decline rate of the light curve (Fig. \ref{fig:lightcurve}), both of which are signs of ejecta-CSM interaction. This also aligns with the beginning of the blueshift visible in the H$\alpha$ line profiles ($\sim-$350 \kms at +230\,d, Fig. \ref{fig:halpha_velocity}), as the ejecta-CSM interaction and CDS are being exposed, having previously been hidden below the CSM.

A natural explanation for blueshifted emission lines is the condensation of new dust grains that increasingly block the receding part of the expanding ejecta \citep{Lucy1989}. Indeed, \cite{Smith2020} interpreted the increasing blueshift of the emission lines from +24\,d to +848\,d in SN~2017hcc as resulting from newly formed dust in either a post-shock CDS or in the SN ejecta.

However, upon closer examination of the centroid of the intermediate component of H$\alpha$ over time (Fig. \ref{fig:halpha_centre}), we find a maximum blueshift ($\sim$7\,{\AA}) at the start of season two. After this point the peak evolves redwards, almost to rest until around +430\,d and after this epoch we see a gradual evolution towards the blue again, but not as large as that seen at earlier epochs. In the final four epochs, +1173\,d to +1765\,d we see a shift back towards the rest wavelength.

It is likely that the early blueshifted emission results from the fast-moving (a few 1000\,\kms) dense shell of material consisting of decelerated ejecta and swept-up CSM. This, combined with the fact that it is also optically thick in the continuum, means that line photons emitted in the red are preferentially absorbed \citep{hillier2012, Dessart2015}.
Indeed, the decreasing blueshift together with a lack of a sudden and marked increase in the infrared flux argues against dust formation in the CDS at these epochs. 
However, as noted above, the situation is complicated by an excursion bluewards between $\sim$500-1000\,d, possibly indicative of the onset of grain condensation, facilitated by the decreasing optical depth of the shell, allowing clumpy material therein to cool down. However, taking the latest spectra ($\gtrsim$1000\,d) at face value, any condensation of grains appears to occur in a short and finite window.
In light of the above, it therefore seems likely that the IR emission in SN~2017hcc originated from a combination of pre-existing and newly formed dust similar to a number of other IIn SNe, for example SN~2010jl \citep{Chugai2018}.

By +399\,d, +773\,d, and +1020\,d our two component black body fits yield a T = 1640, 1160, and 940\,K for the IR component, respectively. Within the errors, the black body radius remains constant until +1020\,d with the black body velocity at this epoch corresponding to $\sim$2600~{\kms}. This is consistent with the dust being located in the CDS or in the SN ejecta. 

Previously, dust formation in the post-shock CDS has been observed, such as in SN~1998S, where the dust formation was suggested to have taken place behind the reverse shock \citep{Pozzo2004}, and in the SN Ibn SN~2006jc, where the dust formation probably occurred behind the forward shock \citep{Mattila2008}. It is difficult to pinpoint the exact location given that different, and time dependent processes both destroy and form dust grains as the SN evolves.

SN~2017hcc shows particular similarities to SN~2010jl, especially in terms of overall brightness and spectral evolution. Both showed blueshifted H$\alpha$ profiles up to late times, with SN~2010jl having such a profile at least as late as +1287\,d \citep{Bevan2020}, though the blueshift may be fading in the case of SN~2017hcc at a comparable epoch. In the case of SN~2010jl the blueshift was explained by \cite{Fransson2014} as resulting from a highly asymmetric bulk motion of the scattering medium towards the observer. A similar blueshift in the H$\alpha$ was also observed in SN~2015da and was explained in the same fashion \citep{Tartaglia2020}. 
Interestingly, SN~2017hcc shows a symmetric H$\alpha$ profile around +1400\,d, whereas SN~2010jl displays a clear multi-component structure at a comparable epoch.
Combined with the significant (few percent) polarisation measurements \citep{PolarisationATel2017, Patat2011}, this possibly implies a similar configuration of the CSM in these objects, but with different viewing angles to the line of sight.

Long-lived SNe IIn tend to be found in low metallicity hosts, comparable to the LMC, making SN~2017hcc an apparent outlier in this respect. SN~2010jl, for example, was found in a low metallicity irregular dwarf host with metallicity 12+log(O/H) = 8.2 $\pm$ 0.1 dex and quite possibly had a low metallicity progenitor \citep{Stoll2011}. However, the existence of at least some SNe IIn with a very massive H-rich CSM in near-solar metallicity environments (e.g. 8.48 dex for SN~2015da, \citealt{Tartaglia2020}) complicates this picture. 
It is clear that objects such as SN~2017hcc require careful and detailed modelling to disentangle the progenitor and explosion properties from those of the CSM. It is also clear that results and trends from other studies of long-lived type IIn SNe can only be applied in a qualitative sense, and it remains to be determined whether these SNe share a common combination of parameters that determine their behaviour.

\section*{Acknowledgements}
\begin{acknowledgements}
We thank the anonymous referee for carefully reading the manuscript.
SM acknowledges support from the Magnus Ehrnrooth Foundation and the Vilho, Yrj\"{o}, and Kalle V\"{a}is\"{a}l\"{a} Foundation.
MF is supported by a Royal Society – Science Foundation Ireland University Research Fellowship. 
RK acknowledges support from the Academy of Finland. 
SJB acknowledges support from Science Foundation Ireland and the Royal Society (RS-EA/3471). 
HK was funded by the Academy of Finland projects 324504 and 328898. 
AR acknowledges support from ANID BECAS/DOCTORADO NACIONAL 21202412. 
GL was supported by a research grant (19054) from VILLUM FONDEN. 
LT acknowledges support from MIUR (PRIN 2017 grant 20179ZF5KS). 
LG acknowledges financial support from the Spanish Ministerio de Ciencia e Innovaci\'on (MCIN), the Agencia Estatal de Investigaci\'on (AEI) 10.13039/501100011033, and the European Social Fund (ESF) "Investing in your future" under the 2019 Ram\'on y Cajal programme RYC2019-027683-I and the PID2020-115253GA-I00 HOSTFLOWS project, from Centro Superior de Investigaciones Cient\'ificas (CSIC) under the PIE project 20215AT016, and the programme Unidad de Excelencia Mar\'ia de Maeztu CEX2020-001058-M. 
MN is supported by the European Research Council (ERC) under the European Union's Horizon 2020 research and innovation programme (grant agreement No.~948381) and by a Fellowship from the Alan Turing Institute. 
TMR acknowledges the financial support of the Jenny and Antti Wihuri foundation and the Finnish Academy of Science and Letters through the Finnish postdoc pool and the Vilho, Yrj{\"o} and Kalle V{\"a}is{\"a}l{\"a} Foundation. 
NER acknowledges partial support from MIUR, PRIN 2017 (grant 20179ZF5KS), from the Spanish MICINN grant PID2019-108709GB-I00 and FEDER funds, and from the programme Unidad de Excelencia María de Maeztu CEX2020-001058-M. 
MG is supported by the EU Horizon 2020 research and innovation programme under grant agreement No 101004719. 
TEMB acknowledges financial support from the Spanish Ministerio de Ciencia e Innovaci\'on (MCIN), the Agencia Estatal de Investigaci\'on (AEI) 10.13039/501100011033 under the PID2020-115253GA-I00 HOSTFLOWS project, and from Centro Superior de Investigaciones Cient\'ificas (CSIC) under the PIE project 20215AT016 and the I-LINK 2021 LINKA20409. TEMB was also partially supported by the programme Unidad de Excelencia Mar\'ia de Maeztu CEX2020-001058-M. 
MS acknowledges the Infrared Telescope Facility, which is operated by the University of Hawaii under contract 80HQTR19D0030 with the National Aeronautics and Space Administration. 
Based on observations made with the Nordic Optical Telescope, owned in collaboration by the University of Turku and Aarhus University, and operated jointly by Aarhus University, the University of Turku and the University of Oslo, representing Denmark, Finland and Norway, the University of Iceland and Stockholm University at the Observatorio del Roque de los Muchachos, La Palma, Spain, of the Instituto de Astrofisica de Canarias. The data presented here were obtained in part with ALFOSC, which is provided by the Instituto de Astrofisica de Andalucia (IAA) under a joint agreement with the University of Copenhagen and NOT. 
This work makes use of observations from the LCOGT network. Based on observations collected at Copernico and Schmidt telescopes (Asiago, Italy) of the INAF – Osservatorio Astronomico di Padova, and the 1.22 Galileo Galilei Telescope of the Padova University in the Asiago site. 

We acknowledge ESA Gaia, DPAC and the Photometric Science Alerts Team\footnote{\url{http://gsaweb.ast.cam.ac.uk/alerts}}. 
Based on observations made with the Gran Telescopio Canarias (GTC), installed at the Spanish Observatorio del Roque de los Muchachos of the Instituto de Astrofísica de Canarias, on the island of La Palma. This work is (partly) based on data obtained with the instrument OSIRIS, built by a Consortium led by the Instituto de Astrofísica de Canarias in collaboration with the Instituto de Astronomía of the Universidad Autónoma de México. OSIRIS was funded by GRANTECAN and the National Plan of Astronomy and Astrophysics of the Spanish Government.
The Liverpool Telescope is operated on the island of La Palma by Liverpool John Moores University in the Spanish Observatorio del Roque de los Muchachos of the Instituto de Astrofisica de Canarias with financial support from the UK Science and Technology Facilities Council.
Based on observations made with the Italian Telescopio Nazionale Galileo (TNG) operated on the island of La Palma by the Fundación Galileo Galilei of the INAF (Istituto Nazionale di Astrofisica) at the Spanish Observatorio del Roque de los Muchachos of the Instituto de Astrofisica de Canarias.
This publication makes use of data products from the Two Micron All Sky Survey, which is a joint project of the University of Massachusetts and the Infrared Processing and Analysis Center/California Institute of Technology, funded by the National Aeronautics and Space Administration and the National Science Foundation.
This research has made use of the APASS database, located at the AAVSO web site. Funding for APASS has been provided by the Robert Martin Ayers Sciences Fund. 
Based on observations collected at the European Southern Observatory under ESO programmes 199.D-0143, 1101.D-0105, and 1103.D-0328. 
Based on observations collected at the European Organisation for Astronomical Research in the Southern Hemisphere under ESO programme 0103.D-0440.

Based on data obtained from the ESO Science Archive Facility.

The Pan-STARRS1 Surveys (PS1) have been made possible through contributions of the Institute for Astronomy, the University of Hawaii, the Pan-STARRS Project Office, the Max-Planck Society and its participating institutes, the Max Planck Institute for Astronomy, Heidelberg, and the Max Planck Institute for Extraterrestrial Physics, Garching, The Johns Hopkins University, Durham University, the University of Edinburgh, Queen's University Belfast, the Harvard-Smithsonian Center for Astrophysics, the Las Cumbres Observatory Global Telescope Network Incorporated, the National Central University of Taiwan, the Space Telescope Science Institute, the National Aeronautics and Space Administration Grants No.s NNX08AR22G, NNX12AR65G, and NNX14AM74G, the National Science Foundation under Grant No. AST-1238877, the University of Maryland, Eotvos Lorand University (ELTE), the Los Alamos National Laboratory and the Gordon and Betty Moore Foundation.

This work has made use of data from the Asteroid Terrestrial-impact Last Alert System (ATLAS) project. ATLAS is primarily funded to search for near earth asteroids through NASA grants NN12AR55G, 80NSSC18K0284, and 80NSSC18K1575; byproducts of the NEO search include images and catalogues from the survey area. The ATLAS science products have been made possible through the contributions of the University of Hawaii Institute for Astronomy, the Queen's University Belfast, the Space Telescope Science Institute, and the South African Astronomical Observatory. 

This publication also makes use of data products from NEOWISE, which is a project of the Jet Propulsion Laboratory/California Institute of Technology, funded by the Planetary Science Division of the National Aeronautics and Space Administration. 

\end{acknowledgements}
\bibliographystyle{aa} 
\bibliography{mybib}
\theendnotes
\newpage
\begin{appendix}
\section{Tables}
\input{tables/photometry_log.tex}

\input{tables/spectra_log.tex}
\input{tables/comp_objects.tex}

\end{appendix}
\end{document}

%% file: tables/photometry_log.tex
\onecolumn 
\begin{landscape}
\begin{longtable}{cccccccccc}

\caption{Optical photometry, apart from that of {\it Swift}. The $B$ and $V$ filters are in Vega magnitudes and the {\it griz} filters are in AB magnitudes.}

\label{tab:optphot}\\

\hline
\hline
       Date &      MJD &   Epoch &       $B$ (err) &       $V$ (err) &       $g$ (err) &       $r$ (err) &       $i$ (err) &       $z$ (err) &         Telescope (Instrument) \\
       &       &   (d) &     &        &       &    &    &      &      \\

\hline
 2017-11-03 &  58060.9 &    33.5 &  13.85 (0.02) &  13.75 (0.03) &  14.04 (0.01) &  13.81 (0.05) &  14.05 (0.05) &  14.34 (0.07) &                      LT (IO:O) \\
 2017-11-10 &  58067.0 &    39.6 &     - &     - &     - &     - &     - &  14.30 (0.94) &       NOT (ALFOSC) \\
 2017-11-14 &  58071.9 &    44.5 &  14.02 (0.05) &  13.60 (0.03) &  13.60 (0.06) &  13.64 (0.04) &  13.81 (0.05) &  13.99 (0.04) &                      LT (IO:O) \\
 2017-11-16 &  58073.9 &    46.5 &  13.79 (0.06) &  13.62 (0.04) &  13.61 (0.03) &  13.63 (0.04) &  13.77 (0.05) &  13.98 (0.06) &                      LT (IO:O) \\
 2017-11-17 &  58074.8 &    47.4 &  13.52 (0.09) &  13.66 (0.08) &  13.64 (0.06) &  13.70 (0.07) &  13.78 (0.05) &     - &  Asiago Schmidt (Moravian) \\
 2017-11-18 &  58075.0 &    47.6 &  13.76 (0.03) &  13.62 (0.06) &  13.63 (0.05) &  13.71 (0.04) &  13.71 (0.05) &  13.95 (0.05) &       NOT (ALFOSC) \\
 2017-11-19 &  58076.9 &    49.5 &     - &  13.60 (0.05) &  13.60 (0.08) &  13.56 (0.06) &  13.82 (0.08) &  13.95 (0.04) &                      LT (IO:O) \\
 2017-11-20 &  58077.9 &    50.5 &  13.85 (0.06) &  13.63 (0.05) &  13.63 (0.05) &  12.46 (0.06) &  13.72 (0.02) &     - &       NOT (ALFOSC) \\
 2017-11-26 &  58083.8 &    56.4 &  13.80 (0.03) &     - &     - &     - &     - &     - &  Asiago Schmidt (Moravian) \\
 2017-11-26 &  58083.9 &    56.5 &     - &  13.65 (0.02) &  13.64 (0.02) &  13.64 (0.03) &  13.74 (0.03) &     - &  Asiago Schmidt (Moravian) \\
 2017-11-30 &  58087.9 &    60.5 &  13.91 (0.06) &  13.67 (0.07) &  13.70 (0.06) &  13.68 (0.03) &  13.66 (0.09) &  13.87 (0.08) &       NOT (ALFOSC) \\
 2017-12-03 &  58090.8 &    63.4 &  13.90 (0.04) &  13.70 (0.05) &  13.74 (0.03) &  13.67 (0.03) &  13.75 (0.03) &     - &  Asiago Schmidt (Moravian) \\
 2017-12-05 &  58092.8 &    65.4 &  13.92 (0.03) &  13.70 (0.03) &  13.72 (0.02) &  13.66 (0.02) &  13.76 (0.03) &     - &  Asiago Schmidt (Moravian) \\
 2017-12-07 &  58094.0 &    66.6 &     - &     - &     - &  13.72 (0.02) &  13.26 (0.01) &     - &       NOT (ALFOSC) \\
 2017-12-09 &  58096.7 &    69.3 &     - &  13.74 (0.03) &  13.76 (0.02) &  13.70 (0.02) &  13.77 (0.03) &     - &  Asiago Schmidt (Moravian) \\
 2017-12-09 &  58096.8 &    69.4 &  13.98 (0.03) &     - &     - &     - &     - &     - &  Asiago Schmidt (Moravian) \\
 2017-12-18 &  58105.7 &    78.3 &  14.22 (0.06) &     - &  13.97 (0.05) &     - &     - &     - &  Asiago Schmidt (Moravian) \\
 2017-12-18 &  58105.8 &    78.4 &     - &   13.90 (0.07) &     - &  13.81 (0.06) &  13.86 (0.05) &     - &  Asiago Schmidt (Moravian) \\
 2017-12-23 &  58110.9 &    83.5 &  14.31 (0.03) &  13.84 (0.13) &  14.03 (0.04) &  13.79 (0.03) &  13.81 (0.07) &  14.00 (0.07) &       NOT (ALFOSC) \\
 2017-12-29 &  58116.8 &    89.4 &  14.52 (0.15) &  13.99 (0.15) &  14.24 (0.09) &  13.98 (0.09) &   13.88 (0.1) &  14.12 (0.04) &       NOT (ALFOSC) \\
 2018-01-10 &  58128.7 &   101.3 &  14.90 (0.06) &  14.38 (0.05) &  14.54 (0.05) &  14.21 (0.05) &  14.19 (0.05) &     - &  Asiago Schmidt (Moravian) \\
 2018-01-12 &  58130.8 &   103.4 &  14.98 (0.13) &  14.31 (0.15) &  14.68 (0.06) &  14.27 (0.04) &  14.19 (0.02) &  14.29 (0.04) &       NOT (ALFOSC) \\
 2018-01-19 &  58137.8 &   110.4 &     - &  14.49 (0.18) &  14.79 (0.07) &  14.39 (0.12) &   14.40 (0.03) &     - &       NOT (ALFOSC) \\
 2018-05-19 &  58257.4 &   230.0 &  15.84 (0.02) &  16.16 (0.03) &  16.26 (0.03) &  15.54 (0.03) &  15.99 (0.04) &   15.50 (0.05) &                     LCO (fl03) \\
 2018-05-22 &  58260.4 &   233.0 &  16.78 (0.03) &  16.22 (0.04) &  16.29 (0.04) &  15.58 (0.04) &  16.05 (0.05) &  15.52 (0.06) &                     LCO (fl03) \\
 2018-05-28 &  58266.1 &   238.7 &   16.82 (0.1) &     - &     - &  15.61 (0.09) &  16.11 (0.05) &  15.50 (0.05) &                     LCO (fl16) \\
 2018-06-02 &  58271.4 &   244.0 &  16.84 (0.04) &  16.31 (0.04) &  16.41 (0.02) &  15.68 (0.03) &  16.18 (0.04) &  15.64 (0.05) &                     LCO (fl15) \\
 2018-06-11 &  58280.1 &   252.7 &  16.91 (0.03) &  16.33 (0.02) &  16.46 (0.01) &  15.68 (0.03) &  16.13 (0.04) &  15.76 (0.08) &                     LCO (fl06) \\
 2018-06-17 &  58286.3 &   258.9 &  16.92 (0.05) &  16.21 (0.02) &  16.54 (0.05) &  15.70 (0.04) &  16.17 (0.06) &  15.71 (0.05) &                     LCO (fl03) \\
 2018-06-22 &  58291.3 &   263.9 &  16.99 (0.03) &  16.42 (0.05) &  16.55 (0.02) &  15.82 (0.04) &  16.26 (0.05) &     - &                     LCO (fl15) \\
 2018-06-27 &  58296.4 &   269.0 &  17.00 (0.05) &  16.51 (0.05) &  16.63 (0.04) &  15.82 (0.04) &  16.33 (0.06) &     - &                     LCO (fl15) \\
 2018-07-07 &  58306.1 &   278.7 &  17.08 (0.04) &  16.61 (0.04) &  16.73 (0.03) &  15.90 (0.03) &     - &     - &                     LCO (fl06) \\
 2018-07-07 &  58306.2 &   278.8 &     - &     - &     - &     - &  16.44 (0.05) &     - &                     LCO (fl06) \\
 2018-07-11 &  58310.3 &   282.9 &  17.17 (0.03) &  16.65 (0.03) &  16.73 (0.02) &  15.92 (0.02) &  16.47 (0.04) &     - &                     LCO (fl03) \\
 2018-07-14 &  58313.8 &   286.4 &  17.17 (0.02) &  16.70 (0.03) &  16.79 (0.02) &  16.00 (0.03) &  16.55 (0.06) &     - &                     LCO (fl11) \\
 2018-07-20 &  58319.8 &   292.4 &  17.20 (0.03) &  16.75 (0.03) &  16.83 (0.02) &  16.05 (0.02) &  16.58 (0.04) &     - &                     LCO (fl12) \\
 2018-07-22 &  58321.8 &   294.4 &  17.29 (0.04) &  16.79 (0.04) &  16.87 (0.02) &  16.09 (0.03) &  16.62 (0.05) &     - &                     LCO (fl12) \\
 2018-07-23 &  58322.6 &   295.2 &  17.19 (0.06) &  16.77 (0.05) &  16.89 (0.03) &  16.08 (0.03) &  16.61 (0.06) &     - &                     LCO (fl12) \\
 2018-07-25 &  58324.8 &   297.4 &  17.31 (0.03) &  16.81 (0.03) &  16.92 (0.02) &  16.13 (0.02) &  16.67 (0.05) &     - &                     LCO (fl12) \\
 2018-07-29 &  58328.7 &   301.3 &  17.35 (0.04) &  16.86 (0.05) &  16.95 (0.03) &  16.19 (0.03) &  16.69 (0.05) &     - &                     LCO (fl12) \\
 2018-08-04 &  58334.8 &   307.4 &  17.43 (0.03) &  16.95 (0.04) &  17.03 (0.03) &  16.24 (0.03) &  16.83 (0.06) &     - &                     LCO (fl11) \\
 2018-08-11 &  58341.3 &   313.9 &  17.50 (0.05) &  17.04 (0.08) &  17.10 (0.05) &  16.35 (0.07) &  17.03 (0.08) &     - &                     LCO (fl03) \\
 2018-08-18 &  58348.1 &   320.7 &  17.49 (0.04) &  17.13 (0.04) &  17.12 (0.03) &  16.45 (0.03) &  17.10 (0.04) &     - &                     LCO (fl06) \\
 \hline
 \newpage
 \caption{Continued.}
 \\
 \hline
 \hline
Date &      MJD &   Epoch &       $B$ (err) &       $V$ (err) &       $g$ (err) &       $r$ (err) &       $i$ (err) &       $z$ (err) &         Telescope (Instrument) \\
       &       &   (d) &     &        &       &    &    &      &      \\
 \hline
 2018-08-21 &  58351.5 &   324.1 &  17.64 (0.04) &  17.21 (0.05) &  17.18 (0.03) &  16.46 (0.03) &  17.18 (0.04) &     - &                     LCO (fl12) \\
 2018-09-01 &  58362.3 &   334.9 &  17.83 (0.06) &  17.39 (0.06) &  17.35 (0.05) &  16.61 (0.04) &  17.37 (0.06) &     - &                     LCO (fl03) \\
 2018-09-14 &  58375.7 &   348.3 &  17.89 (0.03) &  17.54 (0.04) &  17.47 (0.03) &  16.83 (0.03) &  17.56 (0.04) &     - &                     LCO (fl12) \\
 2018-09-28 &  58389.8 &   362.4 &  18.02 (0.04) &  17.63 (0.05) &  17.63 (0.03) &  17.02 (0.04) &  17.69 (0.05) &     - &                     LCO (fl16) \\
 2018-10-01 &  58392.1 &   364.7 &  18.20 (0.04) &  17.76 (0.05) &  17.83 (0.05) &  17.21 (0.06) &  17.55 (0.05) &  17.09 (0.04) &       NOT (ALFOSC) \\
 2018-10-14 &  58405.9 &   378.5 &  18.17 (0.05) &  17.88 (0.08) &  17.80 (0.05) &  17.19 (0.04) &  17.87 (0.07) &     - &                     LCO (fl16) \\
 2018-10-15 &  58406.0 &   378.6 &     - &     - &     - &     - &  17.89 (0.06) &     - &                     LCO (fl16) \\
 2018-10-26 &  58417.2 &   389.8 &   18.30 (0.12) &  18.05 (0.09) &  18.03 (0.06) &   17.40 (0.04) &  17.99 (0.08) &     - &                     LCO (fa03) \\
 2018-10-29 &  58420.2 &   392.8 &     - &   17.90 (0.08) &  17.83 (0.04) &   17.33 (0.1) &  16.84 (0.77) &     - &                     LCO (fa03) \\
 2018-10-31 &  58422.2 &   394.8 &   18.54 (0.1) &  18.15 (0.08) &  18.02 (0.04) &  17.46 (0.04) &  18.16 (0.06) &     - &                     LCO (fa03) \\
 2018-11-04 &  58426.0 &   398.6 &  18.38 (0.07) &  18.06 (0.06) &  18.03 (0.04) &  17.49 (0.05) &  18.13 (0.08) &     - &                     LCO (fa15) \\
 2018-11-05 &  58427.4 &   400.0 &  18.24 (0.31) &  18.24 (0.24) &  17.94 (0.15) &  16.63 (0.09) &  18.11 (0.59) &     - &                     LCO (fl11) \\
 2018-11-16 &  58438.8 &   411.4 &  18.44 (0.17) &     - &     - &     - &     - &     - &                     LCO (fa06) \\
 2018-11-16 &  58438.9 &   411.5 &     - &  18.25 (0.17) &  18.30 (0.13) &  17.74 (0.09) &  18.15 (0.18) &     - &                     LCO (fa06) \\
 2018-11-20 &  58442.9 &   415.5 &  18.74 (0.06) &  18.37 (0.07) &  18.36 (0.07) &  17.96 (0.04) &  18.32 (0.03) &  17.67 (0.05) &       NOT (ALFOSC) \\
 2018-11-21 &  58443.9 &   416.5 &  18.49 (0.11) &  17.99 (0.09) &  17.97 (0.06) &  17.64 (0.05) &  18.15 (0.09) &     - &                     LCO (fa06) \\
 2018-11-24 &  58446.1 &   418.7 &  18.69 (0.08) &  18.40 (0.06) &  18.29 (0.05) &     - &     - &     - &                     LCO (fa03) \\
 2018-11-24 &  58446.2 &   418.8 &     - &     - &     - &  17.77 (0.05) &  18.45 (0.07) &     - &                     LCO (fa03) \\
 2018-11-27 &  58449.5 &   422.1 &   18.50 (0.2) &  18.30 (0.05) &  18.32 (0.03) &  17.87 (0.03) &  18.49 (0.06) &     - &                     LCO (fl12) \\
 2018-12-10 &  58462.1 &   434.7 &  18.86 (0.03) &  18.55 (0.05) &  18.47 (0.03) &  17.99 (0.03) &  18.58 (0.07) &     - &                     LCO (fa03) \\
 2018-12-21 &  58473.1 &   445.7 &   18.97 (0.1) &  18.50 (0.08) &  18.55 (0.07) &  18.03 (0.05) &  18.62 (0.1) &     - &                     LCO (fa05) \\
 2018-12-28 &  58480.9 &   453.5 &  18.97 (0.07) &  18.64 (0.07) &  18.67 (0.06) &  18.30 (0.07) &  18.54 (0.09) &   18.19 (0.1) &       NOT (ALFOSC) \\
 2019-06-05 &  58639.4 &   612.0 &   19.71 (0.1) &  19.67 (0.12) &  19.50 (0.11) &  19.31 (0.12) &  19.76 (0.37) &     - &                     LCO (fa03) \\
 2019-06-13 &  58647.2 &   619.8 &  19.83 (0.25) &  19.76 (0.28) &     - &     - &     - &     - &       NOT (ALFOSC) \\
 2019-06-20 &  58654.2 &   626.8 &  20.08 (0.16) &  19.87 (0.07) &  19.86 (0.04) &  19.72 (0.05) &  19.97 (0.12) &   19.74 (0.1) &       NOT (ALFOSC) \\
 2019-06-22 &  58656.8 &   629.4 &  20.06 (0.18) &  19.77 (0.17) &  19.37 (0.16) &  19.59 (0.16) &  19.25 (0.31) &     - &                     LCO (fa11) \\
 2019-07-12 &  58676.4 &   649.0 &  19.90 (0.08) &   19.44 (0.1) &  19.49 (0.08) &  19.28 (0.15) &  16.62 (0.76) &     - &                     LCO (fa05) \\
 2019-07-21 &  58685.2 &   657.8 &   20.25 (0.1) &  19.97 (0.06) &  20.04 (0.08) &  19.84 (0.05) &  20.11 (0.13) &   20.03 (0.1) &       NOT (ALFOSC) \\
 2019-07-24 &  58688.8 &   661.4 &  19.95 (0.11) &  19.67 (0.12) &  19.92 (0.07) &  19.75 (0.08) &  20.14 (0.2) &     - &                     LCO (fa12) \\
 2019-08-04 &  58699.8 &   672.4 &  20.24 (0.06) &  20.15 (0.07) &  20.06 (0.05) &  19.89 (0.07) &  20.17 (0.15) &     - &                     LCO (fa11) \\
 2019-08-05 &  58700.1 &   672.7 &  20.17 (0.29) &  19.66 (0.18) &  19.90 (0.06) &  19.74 (0.09) &  19.94 (0.12) &  20.05 (0.12) &       NOT (ALFOSC) \\
 2019-08-13 &  58708.0 &   680.6 &  19.98 (0.19) &  19.55 (0.19) &  19.88 (0.12) &  19.26 (0.11) &  19.71 (0.13) &  19.53 (0.19) &       NOT (ALFOSC) \\
 2019-08-17 &  58712.1 &   684.7 &  20.29 (0.27) &  19.89 (0.28) &  20.04 (0.19) &  19.78 (0.23) &  20.14 (0.34) &  20.39 (0.8) &       NOT (ALFOSC) \\
 2019-08-20 &  58715.2 &   687.8 &  19.96 (0.17) &  19.96 (0.13) &  19.87 (0.12) &  19.68 (0.15) &  20.06 (0.18) &  20.23 (0.33) &       NOT (ALFOSC) \\
 2019-08-20 &  58715.4 &   688.0 &  20.32 (0.33) &  20.04 (0.27) &  20.09 (0.23) &  19.68 (0.18) &  20.09 (0.34) &     - &                     LCO (fa05) \\
 2019-09-02 &  58728.3 &   700.9 &  20.29 (0.06) &  20.07 (0.07) &  20.01 (0.06) &     - &     - &     - &                     LCO (fa15) \\
 2019-09-02 &  58728.4 &   701.0 &     - &     - &     - &  19.98 (0.07) &   20.4 (0.15) &     - &                     LCO (fa15) \\
 2019-09-09 &  58735.0 &   707.6 &  19.91 (0.17) &  19.26 (0.09) &  19.78 (0.11) &   18.89 (0.1) &  19.16 (0.11) &  19.09 (0.17) &       NOT (ALFOSC) \\
 2019-09-16 &  58742.0 &   714.6 &  20.37 (0.16) &  20.25 (0.14) &  20.17 (0.13) &  20.01 (0.15) &  20.10 (0.28) &  20.58 (0.3) &       NOT (ALFOSC) \\
 2019-09-17 &  58743.2 &   715.8 &  20.70 (0.55) &     - &     - &     - &     - &     - &                     LCO (fa05) \\
 2019-09-17 &  58743.3 &   715.9 &  20.38 (0.44) &  20.02 (0.33) &  19.83 (0.23) &  20.05 (0.48) &   21.17 (1.6) &     - &                     LCO (fa05) \\
 2019-09-28 &  58754.3 &   726.9 &   20.39 (0.1) &  20.14 (0.11) &  20.06 (0.07) &  19.88 (0.09) &   20.58 (0.2) &     - &                     LCO (fa03) \\
 2019-10-09 &  58765.0 &   737.6 &  20.14 (0.35) &  19.93 (0.67) &  20.57 (1.44) &  19.69 (0.75) &   19.45 (0.6) &     - &                     LCO (fa15) \\
 \hline
 \newpage
 \caption{Continued.}
 \\
 \hline
 \hline
 Date &      MJD &   Epoch &       $B$ (err) &       $V$ (err) &       $g$ (err) &       $r$ (err) &       $i$ (err) &       $z$ (err) &         Telescope (Instrument) \\
       &       &   (d) &     &        &       &    &    &      &      \\
 \hline
 2019-10-20 &  58776.2 &   748.8 &   19.89 (0.1) &   19.36 (0.1) &  19.65 (0.11) &  19.63 (0.13) &     - &     - &                     LCO (fa15) \\
 2019-10-20 &  58776.3 &   748.9 &     - &     - &     - &     - &  19.65 (0.16) &     - &                     LCO (fa15) \\
 2019-10-31 &  58787.0 &   759.6 &   19.97 (0.1) &     - &     - &     - &     - &     - &                     LCO (fa03) \\
 2019-10-31 &  58787.1 &   759.7 &     - &  19.82 (0.13) &  19.92 (0.09) &  20.02 (0.09) &  20.22 (0.16) &     - &                     LCO (fa03) \\
 2019-11-04 &  58791.0 &   763.6 &  20.20 (0.15) &  19.98 (0.15) &  20.33 (0.06) &  20.09 (0.09) &  19.92 (0.13) &  20.02 (0.24) &       NOT (ALFOSC) \\
 2019-11-11 &  58798.1 &   770.7 &  20.12 (0.31) &     - &     - &     - &     - &     - &                     LCO (fa03) \\
 2019-11-11 &  58798.2 &   770.8 &     - &   19.55 (0.2) &  19.72 (0.18) &  19.42 (0.16) &  20.29 (0.35) &     - &                     LCO (fa03) \\
 2019-11-17 &  58804.9 &   777.5 &  20.27 (0.16) &  20.12 (0.15) &  20.29 (0.06) &   19.82 (0.1) &  20.15 (0.11) &  20.26 (0.21) &       NOT (ALFOSC) \\
 2019-11-23 &  58810.0 &   782.6 &  20.29 (0.92) &     - &     - &     - &     - &     - &                     LCO (fa07) \\
 2019-11-23 &  58810.1 &   782.7 &  20.10 (0.22) &  19.41 (0.13) &   19.68 (0.1) &  19.11 (0.15) &  20.01 (0.51) &     - &                     LCO (fa07) \\
 2019-12-01 &  58818.9 &   791.5 &  20.71 (0.06) &  20.42 (0.03) &  20.44 (0.07) &  20.26 (0.08) &  20.00 (0.65) &  20.72 (0.14) &       NOT (ALFOSC) \\
 2019-12-03 &  58820.8 &   793.4 &  20.20 (0.17) &  19.81 (0.17) &     - &     - &     - &     - &                     LCO (fa14) \\
 2019-12-03 &  58820.9 &   793.5 &     - &     - &  19.87 (0.15) &  19.37 (0.13) &  19.69 (0.19) &     - &                     LCO (fa14) \\
 2020-01-15 &  58863.8 &   836.4 &  20.79 (0.11) &  20.54 (0.06) &  20.58 (0.06) &  20.32 (0.15) &  20.43 (0.28) &  20.87 (0.15) &       NOT (ALFOSC) \\
 2020-06-22 &  59022.2 &   994.8 &  20.98 (0.04) &  20.77 (0.06) &  20.65 (0.13) &  20.52 (0.15) &  21.04 (0.11) &  21.40 (0.27) &       NOT (ALFOSC) \\
 2020-07-25 &  59055.1 &  1027.7 &  20.95 (0.07) &  20.62 (0.07) &     - &     - &     - &     - &       NOT (ALFOSC) \\
 2020-07-25 &  59055.2 &  1027.8 &  21.09 (0.07) &  20.82 (0.05) &  20.64 (0.12) &  20.59 (0.09) &  21.06 (0.09) &  21.07 (0.2) &       NOT (ALFOSC) \\
 2020-08-21 &  59082.0 &  1054.6 &  21.02 (0.15) &  20.77 (0.08) &  20.70 (0.09) &  20.62 (0.16) &  21.04 (0.13) &  21.22 (0.24) &       NOT (ALFOSC) \\
 2020-09-14 &  59106.0 &  1078.6 &  21.14 (0.05) &     - &     - &     - &     - &     - &       NOT (ALFOSC) \\
 2020-09-14 &  59106.1 &  1078.7 &  21.15 (0.06) &   20.90 (0.05) &  20.73 (0.16) &  20.53 (0.19) &   21.07 (0.1) &   21.34 (0.2) &       NOT (ALFOSC) \\
 2021-01-18 &  59232.8 &  1205.4 &     - &     - &  20.87 (0.11) &   20.82 (0.1) &  21.18 (0.14) &  21.34 (0.25) &       NOT (ALFOSC) \\
 2021-01-18 &  59232.9 &  1205.5 &     - &     - &     - &     - &  21.22 (0.14) &     - &       NOT (ALFOSC) \\
 2021-07-06 &  59401.2 &  1373.8 &  21.19 (0.15) &  20.90 (0.21) &  20.88 (0.21) &  20.90 (0.23) &   21.40 (0.22) &  21.47 (0.29) &                      LRS (LRS) \\
 2021-10-11 &  59498.1 &  1470.7 &   21.3 (0.05) &  20.89 (0.13) &  20.95 (0.13) &  20.88 (0.12) &  21.32 (0.11) &       - &  NOT (ALFOSC) \\
 2021-11-12 &  59530.9 &  1503.5 &     - &     - &  21.14 (0.13) &     - &     - &       - &    NOT (ALFOSC) \\
 2021-11-13 &  59531.0 &  1503.6 &  21.41 (0.18) &  20.85 (0.18) &     - &  21.01 (0.11) &  21.26 (0.16) &       - &    NOT (ALFOSC) \\
 2022-07-07 &  59767.2 &  1739.8 &     - &     - &     - &  21.16 (0.04) &     - &       - &    NOT (ALFOSC) \\
 2022-07-16 &  59776.2 &  1748.8 &  21.34 (0.27) &  20.85 (0.24) &  21.04 (0.2) &     - &     - &       - &    NOT (ALFOSC) \\
 2022-08-23 &  59814.2 &  1786.8 &  21.41 (0.08) &  20.92 (0.13) &  21.18 (0.11) &     - &   21.40 (0.15) &       - &    NOT (ALFOSC) \\
 2022-09-22 &  59844.0 &  1816.6 &     - &     - &     - &  20.99 (0.11) &    - &       - &    NOT (ALFOSC) \\
 2022-10-20 &  59872.9 &  1845.5 &  21.20 (0.12) &  20.86 (0.11) &  21.03 (0.12) &   20.89 (0.18) &  21.33 (0.15) &  - & NOT (ALFOSC) \\
\hline
\end{longtable}

\end{landscape}
\twocolumn

\begin{table*}
\centering
\caption{{\it Swift} photometry (Vega magnitudes).}
\label{tab:uvphot}
\begin{tabular}{lrrllllll}

\hline
\hline
       Date &      MJD &  Epoch &    $UVW2$ (err) &    $UVM2$ (err) &    $UVW1$ (err) &       $U$ (err) &       $B$ (err) &       $V$ (err) \\
        &    &  (d) &    &   &   &       &       &       \\

\hline
 2017-10-28 &  58054.4 &   27.0 &  12.48 (0.04) &  12.32 (0.04) &   12.40 (0.04) &   12.70 (0.04) &  13.97 (0.04) &  13.93 (0.05) \\
 2017-10-30 &  58056.1 &   28.7 &  12.52 (0.04) &     - &   12.40 (0.04) &  12.68 (0.04) &   13.90 (0.04) &     - \\
 2017-11-04 &  58061.0 &   33.6 &  12.59 (0.04) &  12.39 (0.04) &  12.41 (0.04) &  12.58 (0.04) &   13.80 (0.04) &  13.78 (0.04) \\
 2017-11-05 &  58062.5 &   35.1 &     - &  12.42 (0.04) &     - &     - &     - &     - \\
 2017-11-05 &  58062.8 &   35.4 &  12.62 (0.04) &  12.41 (0.04) &  12.43 (0.04) &  12.58 (0.04) &  13.77 (0.04) &  13.72 (0.04) \\
 2017-11-10 &  58067.7 &   40.3 &  12.78 (0.04) &  12.54 (0.04) &  12.52 (0.04) &  12.60 (0.04) &  13.75 (0.04) &  13.69 (0.04) \\
 2017-11-16 &  58073.5 &   46.1 &  12.97 (0.04) &  12.78 (0.04) &  12.67 (0.04) &  12.66 (0.04) &  13.75 (0.04) &  13.65 (0.04) \\
 2017-11-19 &  58076.6 &   49.2 &  13.12 (0.04) &  12.89 (0.04) &  12.78 (0.04) &  12.69 (0.04) &  13.73 (0.04) &  13.60 (0.04) \\
 2017-11-22 &  58079.6 &   52.2 &  13.27 (0.04) &  13.03 (0.04) &  12.89 (0.04) &  12.72 (0.04) &  13.78 (0.04) &  13.62 (0.04) \\
 2017-11-30 &  58087.6 &   60.2 &  13.76 (0.04) &  13.49 (0.04) &  13.29 (0.04) &  12.94 (0.04) &  13.82 (0.04) &  13.65 (0.04) \\
 2017-12-04 &  58091.6 &   64.2 &  14.02 (0.05) &  13.75 (0.05) &  13.48 (0.04) &  13.10 (0.04) &  13.93 (0.04) &  13.72 (0.04) \\
 2017-12-09 &  58096.1 &   68.7 &  14.33 (0.05) &  14.05 (0.05) &  13.74 (0.04) &  13.24 (0.04) &  13.98 (0.04) &  13.71 (0.04) \\
 2017-12-12 &  58099.3 &   71.9 &  14.55 (0.05) &  14.27 (0.05) &  13.89 (0.05) &  13.38 (0.04) &  14.04 (0.04) &  13.75 (0.04) \\

\hline
\end{tabular}
\label{table:swift}
\end{table*}

\begin{table*}
\caption{Near-infrared photometry.}
\centering
\begin{tabular}{lrrllll}

\hline\hline
       Date &      MJD &   Epoch &       $J$ (err) &       $H$ (err) &      $Ks$ (err) & Telescope (Instrument) \\
        &       &   (d) &       &       &     &  \\

\hline
 2017-11-10 &  58067.2 &    39.8 &  13.08 (0.14) &  13.06 (0.17) &  12.69 (0.09) &        NTT (SOFI) \\
 2017-11-26 &  58083.0 &    55.6 &  12.85 (0.25) &  13.09 (0.14) &  12.71 (0.4) &        NTT (SOFI) \\
 2017-12-14 &  58101.1 &    73.7 &  12.95 (0.15) &  13.05 (0.27) &  12.33 (0.13) &        NTT (SOFI) \\
 2018-01-03 &  58121.9 &    94.5 &  13.25 (0.13) &  13.19 (0.12) &     - &    NOT (NOTCAM) \\
 2018-01-06 &  58124.0 &    96.6 &  13.14 (0.16) &  12.79 (0.24) &  12.34 (0.15) &        NTT (SOFI) \\
 2018-08-19 &  58349.2 &   321.8 &  15.55 (0.14) &  14.83 (0.19) &  13.90 (0.14) &        NTT (SOFI) \\
 2018-09-04 &  58365.1 &   337.7 &  15.62 (0.23) &  14.80 (0.25) &  13.09 (0.14) &    NOT (NOTCAM) \\
 2018-11-03 &  58425.9 &   398.5 &  16.19 (0.19) &  15.03 (0.21) &  13.89 (0.11) &    NOT (NOTCAM) \\
 2019-07-22 &  58686.2 &   658.8 &  18.44 (0.18) &     - &     - &    NOT (NOTCAM) \\
 2019-08-18 &  58713.1 &   685.7 &  18.56 (0.18) &  16.90 (0.21) &  15.23 (0.21) &    NOT (NOTCAM) \\
 2019-10-05 &  58761.0 &   733.6 &  19.09 (0.24) &  17.16 (0.26) &  15.29 (0.15) &    NOT (NOTCAM) \\
 2020-06-25 &  59025.2 &   997.8 &  20.18 (0.33) &  18.65 (0.23) &  16.76 (0.12) &    NOT (NOTCAM) \\
 2020-07-17 &  59047.1 &  1019.7 &  20.61 (0.35) &  18.78 (0.38) &     - &    NOT (NOTCAM) \\
 2020-07-17 &  59047.2 &  1019.8 &     - &     - &   16.85 (0.1) &    NOT (NOTCAM) \\
 2020-12-16 &  59199.9 &  1172.5 &     - &     - &  17.44 (0.14) &    NOT (NOTCAM) \\

\hline
\end{tabular}
\label{table:NIR}
\end{table*}

\begin{table*}
\caption{NEOWISE photometry.}
\centering
\begin{tabular}{lrrllll}

\hline\hline
       Date &      MJD &   Epoch &       $W1$ (err) &       $W2$ (err)  \\
        &       &   (d) &       &       \\

\hline
2017-12-01 & 58088.7 & 61.3 & 12.814 (0.009) & 12.688 (0.020) \\
2018-06-18 & 58287.0 & 259.6 & 13.744 (0.019) & 12.720 (0.034) \\
2018-11-27 & 58449.4 & 422.0 & 13.020 (0.010)	& 12.436 (0.019) \\
2019-06-17 & 58651.2 & 623.8 & 13.362 (0.014)	& 12.584 (0.014) \\
2019-11-26 & 58813.4 & 786.0 & 13.742 (0.022)	& 12.697 (0.026) \\
2020-06-18 & 59018.4 & 991.0 & 14.230 (0.030)	& 13.10	(0.04) \\
2020-11-27 & 59180.6 & 1153.2 & 14.555 (0.023)	& 13.315 (0.032) \\
2021-06-17 & 59382.7 & 1355.3 & 14.93 (0.05) & 13.72 (0.16) \\
2021-11-26 & 59544.9 & 1517.5 & 14.932 (0.015)	& 13.79 (0.04) \\

\hline
\end{tabular}
\label{table:WISE}
\end{table*}

%% file: tables/spectra_log.tex
\begin{table*}[ht]

\caption{Log of optical spectroscopic observations of SN 2017hcc.}

\label{tab:spec}

\centering

\begin{tabular}{lrrrlll}

\hline\hline
       Date &      MJD &  Epoch &  Exposure time & Slit width &   Wavelength range  &              Telescope (instrument, grism) \\
        &       &  (d) & (s) &  &  (Å) &   \\
              \hline

 2017-10-07 &  58033.4 &          6 &            900 &            2.0" &   5000-9500 &           FTS (FLOYDS, red/blue) \\
 2017-11-01 &  58058.9 &       31.5 &            300 &            1.0" &   3200-9600 &                           NOT (ALFOSC, \#4) \\
 2017-11-09 &  58066.2 &       38.8 &            600 &            1.0" &   3380-7520  &                    NTT (EFOSC, Gr\#11) \\
 2017-11-09 &  58066.2 &       38.8 &            600 &            1.0" &   6015-10320 &                     NTT (EFOSC, Gr\#16) \\
 2017-11-27 &  58071.8 &       44.4 &           1800 &            250 $\mu$m &3300-7000  & Asiago 1.22m (B\&C, 300tr) \\
 2017-11-17 &  58075.0 &       47.6 &            420 &            1.3" &   3200-9600      &                      NOT (ALFOSC, \#4) \\
 2017-11-27 &  58083.9 &       56.5 &           1800 &            200 $\mu$m &3300-7000  & Asiago 1.22m (B\&C, 300tr) \\
 2017-11-27 &  58084.1 &       56.7 &            600 &            1.0" &  3380-7520        &               NTT (EFOSC, Gr\#11) \\
 2017-11-27 &  58084.1 &       56.7 &            600 &            1.0" &   6015-10320       &               NTT (EFOSC, Gr\#16) \\
 2017-12-04 &  58090.8 &       63.4 &           1800 &            200 $\mu$m &3300-7000   &Asiago 1.22m (B\&C, 300tr) \\
 2017-12-06 &  58094.0 &       66.6 &            420 &            1.3" &   3200-9600         &                    NOT (ALFOSC, \#4) \\
 2017-12-11 &  58096.8 &       69.4 &           1800 &            200 $\mu$m & 3300-7000  &Asiago 1.22m (B\&C, 300tr) \\
 2017-12-13 &  58100.1 &       72.7 &            600 &            1.0" &  3380-7520       &                NTT (EFOSC, Gr\#11) \\
 2017-12-13 &  58100.1 &       72.7 &            600 &            1.0" &   6015-10320      &                NTT (EFOSC, Gr\#16) \\
 2017-12-21 &  58107.7 &       80.3 &           1200 &            250 $\mu$m &3300-7000   &Asiago 1.22m (B\&C, 300tr) \\
 2017-12-28 &  58115.9 &       88.5 &            600 &            0.9" &  6330-6870        &     NOT (ALFOSC, \#17) \\
 2017-12-29 &  58116.8 &       89.4 &            420 &            1.0" &   3200-9600       &                     NOT (ALFOSC, \#4) \\
 2018-01-14 &  58132.0 &      104.6 &            600 &            1.0" &  3380-7520        &               NTT (EFOSC, Gr\#11) \\
 2018-01-14 &  58132.1 &      104.7 &            600 &            1.0" &  6015-10320         &              NTT (EFOSC, Gr\#16) \\
 2018-05-20 &  58258.4 &      231.0 &            900 &            1.0" &  3380-7520          &             NTT (EFOSC, Gr\#11) \\
 2018-05-20 &  58258.4 &      231.0 &            900 &            1.0" &  6015-10320           &            NTT (EFOSC, Gr\#16) \\
 2018-06-27 &  58296.2 &      268.8 &            900 &            1.0" &  3200-9600             &                NOT (ALFOSC, \#4) \\
 2018-06-30 &  58299.3 &      271.9 &           1500 &            1.0" &  3380-7520       &                NTT (EFOSC, Gr\#11) \\
 2018-06-30 &  58299.3 &      271.9 &           1500 &            1.0" &  6015-10320         &              NTT (EFOSC, Gr\#16) \\
 2018-07-13 &  58312.2 &      284.8 &            900 &            1.3" &  3200-9600       &                      NOT (ALFOSC, \#4) \\
 2018-07-24 &  58323.1 &      295.7 &            900 &            1.0" &  3200-9600         &                    NOT (ALFOSC, \#4) \\
 2018-08-04 &  58334.2 &      306.8 &           1200 &            1.0" &  3200-9600           &                  NOT (ALFOSC, \#4) \\
 2018-08-08 &  58338.2 &      310.8 &           1200 &            1.3" &  3200-9600       &                      NOT (ALFOSC, \#4) \\
 2018-08-11 &  58341.1 &      313.7 &           1800 &            1.69" &  3300-9300        &      Asiago 1.82m (AFOSC, VPH6+VPH7) \\
 2018-08-27 &  58357.1 &      329.7 &           1200 &            1.0" &  3200-9600         &                    NOT (ALFOSC, \#4) \\
 2018-10-01 &  58392.1 &      364.7 &           2000 &            1.0" & 3200-9600           &                   NOT (ALFOSC, \#4) \\
 2018-10-18 &  58409.0 &      381.6 &           1900 &            1.0" & 3200-9600        &                      NOT (ALFOSC, \#4) \\
 2018-11-13 &  58436.0 &      408.6 &           2400 &            1.0" & 3200-9600        &                      NOT (ALFOSC, \#4) \\
 2018-12-04 &  58456.5 &      429.1 &           3600 &            2.0" &  5400-10000        & FTS (FLOYDS, red) \\
 2018-12-06 &  58459.0 &      431.6 &           2400 &            1.3" &  3200-9600       &                      NOT (ALFOSC, \#4) \\
 2018-12-28 &  58480.9 &      453.5 &           2400 &            1.3" &  3200-9600         &                    NOT (ALFOSC, \#4) \\
 2019-05-13 &  58616.4 &      589.0 &           3600 &            1.0" &   3380-7520          &            NTT (EFOSC, Gr\#11) \\
 2019-05-14 &  58617.4 &      590.0 &           3600 &            1.0" &  6015-10320           &            NTT (EFOSC, Gr\#16) \\
 2019-06-13 &  58647.2 &      619.8 &           3600 &            1.0" &  3200-9600            &                 NOT (ALFOSC, \#4) \\
 2019-07-20 &  58684.1 &      656.7 &           3600 &            1.0" &  3200-9600             &                NOT (ALFOSC, \#4) \\
 2019-07-28 &  58692.2 &      664.8 &           2373 &            - &   4700-9400     &  VLT (MUSE, -) \\
 2019-07-30 &  58694.2 &      666.8 &           3380 &            1.0" &  3200-9600           &                  NOT (ALFOSC, \#4) \\
 2019-08-05 &  58700.2 &      672.8 &           3000 &            1.0" &  3200-9600            &                 NOT (ALFOSC, \#4) \\
 2019-08-20 &  58715.2 &      687.8 &           3600 &            1.3" &  3200-9600             &                NOT (ALFOSC, \#4) \\
 2019-11-17 &  58804.9 &      777.5 &           3600 &            1.3" &  3200-9600              &               NOT (ALFOSC, \#4) \\
 2019-12-01 &  58819.0 &      791.6 &           3600 &            1.0" &  3200-9600       &                      NOT (ALFOSC, \#4) \\
 2020-07-02 &  59032.2 &     1004.8 &         1500x2 &            1.0" &  3630-7500 	   &     GTC (OSIRIS, R1000B) \\
 2020-10-12 &  59125.0 &     1107.6 &     1432, 1800 &            1.5" &  3000-8430     &   TNG (DOLORES, LR-B) \\
 2020-12-17 &  59200.8 &     1173.4 &         1800x2 &            1.5" &  3000-8430   &     TNG (DOLORES, LR-B) \\
 2021-07-06 &  59401.1 &     1373.7 &         1800x2 &            1.5" &   3000-8430    &   TNG (DOLORES, LR-B) \\
 2021-09-09 &  59466.0 &     1438.6 &         1800x2 &            1.0" &   3000-8430  &     TNG (DOLORES, LR-B) \\
 2022-08-01 &  59792.2 &     1764.8 &         1800x2 &            1.0" &   5100-10000   &     GTC (OSIRIS, R1000R) \\

\hline
\end{tabular}

\end{table*}

\begin{table*}[ht]
\caption{Log of NIR spectroscopic observations of SN 2017hcc.} 
\label{tab:NIR_spec}
\centering
\begin{tabular}{lrrrlll}

\hline
\hline
       Date &      MJD &  Epoch &  Exposure time & Slit width &   Wavelength range &             Telescope (instrument, grism) \\
            &          &  (d) &  (s) &  &   ($\mu$m) &            \\

\hline        

        2017-11-10 &  58067.2 &   39.8 &  960    &        1.0" &   0.95-1.64 &      NTT (SOFI, GB) \\
        2017-11-10 &  58067.2 &   39.8 &  1500    &        1.0" &   1.53-2.52 &      NTT (SOFI, GR) \\
        2017-11-26 &  58083.1 &   55.7 &  2880    &        1.0" &   0.95-1.64 &      NTT (SOFI, GB) \\
        2017-11-26 &  58083.1 &   55.7 &  4500    &        1.0" &  1.53-2.52 &       NTT (SOFI, GR) \\
        2017-12-11 &  58098.2 &    70.8 &  4191     &         0.5" &  0.8–2.4 &       IRTF (SpeX, ShortXD) \\
        2017-12-14 &  58101.0 &   73.6 &  960    &        1.0" &   0.95-1.64 &      NTT (SOFI, GB) \\ 
        2017-12-14 &  58101.1 &   73.7 &  1500    &       1.0" &   1.53-2.52 &      NTT (SOFI, GR) \\
        2018-09-06 &  58367.3 &    339.9 &  719      &         0.5" &   0.7–2.5 &       IRTF (SpeX, Prism) \\

\hline
\end{tabular}

\end{table*}

%% file: tables/comp_objects.tex
\begin{table*}
\caption{Comparison objects.}
\label{tab:compobjects}

\centering
\begin{tabular}{lrrlp{7.5cm}}

\hline\hline
       Name &      Right ascension &   Declination & Redshift &      Source(s) \\

\hline
 SN~2010jl & 09:42:53.330 & +09:29:41.78 &0.0107 & \citealt{Stoll2011, Patat2011, Smith2011, Andrews2011, Zhang2012, Ofek14, Fransson2014, Borish2015, Jencson16, Li2022, Sarangi2018, Chugai2018, Ofek2019, Bevan2020}  \\
 SN~1998S & 11:46:06.180 & +47:28:55.49& 0.0030& \citealt{Leonard2000, Fassia2000, Fassia2001, Pozzo2004, Fransson2005}\\
 SN~2005ip &09:32:06.420 & +08:26:44.41 & 0.0072 & \citealt{Smith2009, Stritzinger2012} \\
 SN~2015da &13:52:24.110  &+39:41:28.60 &0.0067 & \citealt{Tartaglia2020} \\
 HSC16aayt (SN~2016jiu) &10:02:05.570 & +02:57:58.30 & 0.6814& \citealt{Moriya2019} \\ 
 PTF12glz & 15:54:53.040 & +03:32:07.50 & 0.0799 & \citealt{Soumagnac2019} \\

\hline
\end{tabular}
\end{table*}